\documentclass[conference]{IEEEtran}
\IEEEoverridecommandlockouts
\usepackage{cite}
\usepackage{amsmath,amssymb,amsfonts}
\usepackage{graphicx}
\usepackage{textcomp}
\usepackage{xcolor}
\def\BibTeX{{\rm B\kern-.05em{\sc i\kern-.025em b}\kern-.08em
    T\kern-.1667em\lower.7ex\hbox{E}\kern-.125emX}}

\usepackage{algorithm}
\usepackage{algpseudocode}
\usepackage{comment}

\newtheorem{theorem}{Theorem}
\algdef{SE}[DOWHILE]{Do}{doWhile}{\algorithmicdo}[1]{\algorithmicwhile\ #1}%
\newcommand{\etal}{\textit{et al.}}
\usepackage{rotating}
\usepackage{color} 
\usepackage{tabularx} 
\usepackage{adjustbox} 
\usepackage{multirow} 
\usepackage{cite}
\usepackage{amsmath,amssymb,amsfonts}
\usepackage{textcomp}
\usepackage{xcolor}
\MakeRobust{\Call}

\makeatletter
\newcommand{\linebreakand}{%
  \end{@IEEEauthorhalign}
  \hfill\mbox{}\par
  \mbox{}\hfill\begin{@IEEEauthorhalign}
}
\makeatother

\begin{document}

\title{Secure Remote Attestation with Strong Key Insulation Guarantees\thanks{This research was supported by the National Science Foundation under Grants No. 1617774 and 1929261.} \thanks{This e-print uses material from the e-prints\cite{Dijk2021AutonomousSR,Dijk2021BilinearMB} that were previously published by the same authors to formulate a secure remote attestation protocol.}}

\author{\IEEEauthorblockN{Deniz Gurevin}
\IEEEauthorblockA{
\textit{University of Connecticut, Storrs, CT, USA}\\
deniz.gurevin@uconn.edu}\\
\and
\IEEEauthorblockN{Chenglu Jin}
\IEEEauthorblockA{\textit{CWI Amsterdam, The Netherlands} \\
chenglu.jin@cwi.nl}\\
\and
\IEEEauthorblockN{Phuong Ha Nguyen}
\IEEEauthorblockA{\textit{eBay, San Jose, CA, USA} \\
phuongha.ntu@gmail.com}\\
\and

\linebreakand
\and
\IEEEauthorblockN{Omer Khan}
\IEEEauthorblockA{
\textit{University of Connecticut, Storrs, CT, USA}\\
khan@uconn.edu}\\
\and
\IEEEauthorblockN{Marten van Dijk}
\IEEEauthorblockA{\textit{CWI Amsterdam, The Netherlands} \\
marten.van.dijk@cwi.nl}\\

}

\maketitle
\thispagestyle{plain}
\pagestyle{plain}

\begin{abstract}
Recent years have witnessed a trend of secure processor design in both academia and industry. Secure processors with hardware-enforced isolation can be a solid foundation of cloud computation in the future. However, due to recent side-channel attacks, the commercial secure processors failed to deliver the promises of a secure isolated execution environment. Sensitive information inside the secure execution environment always gets leaked via side channels. This work considers the most powerful software-based side-channel attackers, i.e., an All Digital State Observing (ADSO) adversary who can observe all digital states, including all digital states in secure enclaves. 

Traditional signature schemes are not secure in ADSO adversarial model. We introduce a new cryptographic primitive called One-Time Signature with Secret Key Exposure (OTS-SKE), which ensures no one can forge a valid signature of a new message or nonce even if all secret session keys are leaked. OTS-SKE enables us to sign attestation reports securely under the ADSO adversary. We also minimize the trusted computing base by introducing a secure co-processor into the system, and the interaction between the secure co-processor and the attestation processor is unidirectional. That is, the co-processor takes no inputs from the processor and only generates secret keys for the processor to fetch. Our experimental results show that the signing of OTS-SKE is faster than that of Elliptic Curve Digital Signature Algorithm (ECDSA) used in Intel SGX. 

\end{abstract}


\section{Introduction}

Due to their low cost and practicality, public cloud services are widely used today to employ application services. The advantages of cloud computing include on-demand resource allocation and high availability for services. On the other hand, customers of cloud services have to trust cloud providers, who are in control of the technical infrastructure of cloud services. This includes the hardware and software components that enable the resource sharing of multiple virtual machines (VMs) or enclaves of cloud customers on a single platform, which comes at a risk. Security concerns related to cloud computing lead to the deployment of trusted execution environments (TEE). Several secure processor architectures are widely employed in cloud settings. Intel SGX\cite{Costan2016IntelSE} aims at protecting sensitive data by enabling application isolation technology and protecting enclaves from any process outside the enclave itself, including processes running at higher privilege levels. In August 2020, Intel announced new extensions for its Instruction Set Architecture (ISA), Intel Trust Domain Extensions (Intel TDX) \cite{TDX}, which deploys hardware isolated VMs, called trust domains (TD), and aims to protect them from the virtual-machine manager (VMM)/hypervisor and any other non-TD software. Secure Encrypted Virtualization (SEV) technology by AMD \cite{AMDSEV} also aims to prove the correct and secure deployment of VMs in cloud setting by encrypting the main memory of VMs with VM-specific keys, and preventing higher-privileged hypervisor access to a VM's memory.

Secure processor architecture design\cite{Lie2000ArchitecturalSF, Suh2003AEGISAF, ARM2009, Champagne2010ScalableAS, Boivie2011SecureBlue,  Fletcher2012ASP, Costan2016IntelSE, Costan2016SanctumMH,  Bourgeat2019MI6SE} is based on two core principles: hardware isolation and remote attestation (RA). Hardware isolation allows one to run a code snippet in a so-called enclave that is isolated from the OS and other enclaves with the goal of keeping its internal computations private. Hardware isolation implements access control which, for example, disallows the OS to access reserved DRAM for enclaves. 
Besides being able to execute code in a trusted execution environment that guarantees privacy, a remote user also needs to be able to verify whether a computed result originated from the executed code. Remote attestation is based on digital signature schemes and signs and binds a computed result to the enclave code that produced it and the processor identity. A remote user needs remote attestation to verify the results produced by such an enclave. Usually, an asymmetric key system is adopted for attestation purposes, so that an attestation can be performed using the private key in an isolated platform and then verified from other platforms using the corresponding public key that is known by the verifier. 

On the other hand, hardware isolation, which the remote attestation relies on for its security, has been shown to be elusive. The enclave platform, where the remote attestation is performed, itself may have vulnerabilities and can possibly be exploited through its own I/O interactions. With the developing capabilities of adversaries and side channel attacks, vulnerabilities of secure processors continuously keep getting exploited leaking the private digital state (including private keys) of the processor. A recent survey\cite{Nilsson2020ASO} has shown that Intel SGX has been susceptible to a wide range of attacks in the recent years\cite{iago2013, Lee2017HackingID, Weichbrodt2016AsyncShockES, vanbulck2017sgxstep, vanbulck2019tale, Bulck2018NemesisSM, Gyselinck2018OffLimitsAL, Huo2020BluethunderA2, Wang2017LeakyCO, Bulck2018ForeshadowET, Moghimi2017CacheZoomHS, Gtzfried2017CacheAO, Schwarz2017MalwareGE, Brasser2017SoftwareGE, Dall2018CacheQuoteER, Moghimi2018MemJamAF, Lee2017InferringFC, Evtyushkin2018BranchScopeAN, Kocher2019SpectreAE, Lipp1999Meltdown, Chen2020SgxPectreSI, Schaik2020CacheOutLD, Schaik2019RIDLRI, Schwarz2019ZombieLoadCD, Ragab2021CrossTalkSD, Jang2017SGXBombLD, Kim2014FlippingBI, Zhang2019TeleHammerAF, Murdock2020PlundervoltSF}. We may conclude that \textbf{hardware isolation as is implemented today for executing enclave code cannot guarantee privacy}. In fact, any internally computed enclave value may potentially leak; we cannot make any solid privacy guarantee. 

In this work, we generalize and strengthen the above-mentioned side channel attacks, assume the \textit{worst case} for processor security, and focus on an extremely strong adversarial model, which we call \textbf{All Digital State Observing (ADSO)} adversary. An ADSO adversary can observe all digital states in a processor, including all the intermediate states in enclaves, that may be leaked via known or even currently unknown side channels. Besides, we assume that the integrity of the computation on the processor will not be tampered with by the ADSO adversary, as side channel attacks are generally a passive attack technique. In such a strong ADSO adversarial model, no secrets can be kept safe and confidential computation becomes nearly impossible (unless fully homomorphic encryption is used~\cite{gentry2009fully,van2010fully}). The main research problem we want to solve is \textbf{whether we can still enable verifiable computation, i.e., remote attestation, in an ADSO adversarial model.}

Traditional cryptographic methods do not work anymore under the ADSO adversary because the methods all depend on a secure secret key which does not exist under the ADSO adversary. We propose to introduce a truly isolated piece of hardware that is not affected by any adversary (including the ADSO adversary) -- a secure co-processor that is only dedicated to generating private keys in the used digital signature schemes. Note that one can also introduce an isolated secure co-processor to sign attestation reports for the processor. Still, it requires the co-processor to take inputs from a potentially insecure processor and have bidirectional interactions with the processor. We want to \textit{minimize the attack surface as much as possible} by enforcing a unidirectional interaction between the co-processor and the attestation processor, and hence the co-processor only generates fresh private keys and does not take any inputs.

However, this still means that the secret keys will be handed over to the processor for signing the attestation reports of enclaves, and the keys are exposed to ADSO adversaries. One way of thwarting this problem is discarding and renewing the digital secret keys after each use, effectively breaking the lifetime of a signature scheme into sessions and using session keys. Forward secure schemes~\cite{Itkis2004} and key-insulated schemes (KIS)~\cite{DodisKatzXuEtAl2002,DodisKatzXuEtAl2003} are two typical ways to limit the damage of the session key leakage. Forward secure schemes protect all past session keys when the current session key leaks, while KIS offers a stronger guarantee that the security of any uncompromised sessions is not affected if some session keys are compromised and leaked.

Even key-insulated schemes are still not strong enough for resisting ADSO adversaries because the ADSO adversaries can use the leaked session keys to forge a valid signature specific to that session. To provide complete protection against the strong ADSO adversary, we introduce a new cryptographic notion, \textbf{One-Time Signature with Secret Key Exposure (OTS-SKE)}, which can \textit{guarantee the security of all sessions even if all session keys are exposed to the adversary}. This is because every secret key is unique to the message to be signed and a fresh nonce generated by the signature verifier. 

Relying on the OTS-SKE scheme, we propose a RA protocol between an RA enclave at the processor and a remote user. The one-time secret keys are generated in an isolated environment (co-processor) and sent to the insecure processor via a special memory component, called \textit{oblivious transfer memory}~\cite{goldwasser2008one}, for signing by the RA enclave. The co-processor keeps updating fresh secret keys in the oblivious transfer memory, and each key consists of a group of subkeys. There is a special mechanism implemented in the oblivious transfer memory: after reading a subset of the subkeys chosen by the processor from the oblivious transfer memory, all the subkeys will be overwritten by zeros or erased. Hence, effectively the oblivious transfer memory runs an oblivious transfer with the processor~\cite{kilian1988founding}. All secret keys that once leave the oblivious transfer memory are exposed to the ADSO adversary, and they are used in the signing procedure to sign the attestation report. The signature produced by the RA enclave can be verified using a single universal public key by a remote user. To the best of our knowledge, \textit{OTS-SKE scheme is the only scheme that can secure all sessions/signatures under an ADSO adversary}.

We implemented the proposed OTS-SKE scheme and measured its performance for key generation, signing and verification. Results in Section\ref{sec:performance} show that for a single session, signing takes  $3.4$ ms to produce a signature, and verification by a remote user takes $127.3$ ms. During initialization, generation of all subkeys of a session key takes $388.6$ ms. 

\subsection{Contributions}

\begin{itemize}
    \item We generalize all existing side channel attacks and introduce an extremely powerful adversarial model: All Digital State Observing (ADSO) adversary.
    \item We introduce a novel cryptographic notion: One-Time Signature with Secret Key Exposure (OTS-SKE), which can be secure under ADSO adversaries. Also, we present a concrete construction of OTS-SKE based on bilinear map.
    \item Building upon the OTS-SKE scheme we develop, we propose a remote attestation scheme that can survive in ADSO adversarial model.
    \item We implemented the proposed RA scheme, and the experimental results show that we can reduce the signing phase to $3.4$ ms, compared with Elliptic Curve Digital Signature Algorithm which takes $23.1$ ms. Our verification implementation takes $127.3$ ms and during initialization, generation of all subkeys of a session takes $388.6$ ms.
\end{itemize}

\subsection{Organization}

Section~\ref{sec:RA_background} presents an overview of the current state-of-the-art remote attestation techniques in secure processors. Section~\ref{sec:ADSO} motivates and details the ADSO adversarial model we introduce. Section~\ref{sec:crypto_related_work} compares the existing cryptographic schemes with the proposed OTS-SKE. The definition, construction, and application of OTS-SKE are introduced in Section~\ref{sec:main_scheme}. The implementation details and experimental results are shown in Section~\ref{sec:performance}. Finally, the paper concludes in Section~\ref{sec:conclusion}.

\section{Remote Attestation in Current State-of-the-Art Secure Processor Architecture Technology}\label{sec:RA_background}

Public cloud environments are frequently employed for application services today and they offer cost-effective solutions. In a cloud setting, applications are running in a distributed mode on different servers, where they interact with the outside world as well as with other applications. These applications might contain sensitive code and the interactions with the outside world might lead to malicious access and consequently, private data leakage, which is an undesired situation for cloud clients. Therefore, it is crucial for cloud suppliers to provide secure execution environments and secure communication for the application of its clients.

When a client has a piece of sensitive code (for example, an \textit{enclave} or a \textit{virtual machine (VM)}), she can send it to a cloud provider who provides a secure execution service (e.g. by Intel SGX). The cloud provider must convince the client that her sensitive code is securely running on an authentic Intel SGX processor to earn her trust. This root of trust is accomplished by attestation. \textit{Local attestation} allows an enclave to attest its execution environment to other enclaves on the same platform, while for the processor to authenticate its enclaves' configuration to a remote entity outside the platform, \textit{remote attestation (RA)} needs to be performed. After an enclave attests itself to a remote party, an encrypted communication channel can be established between the two.

Intel SGX currently provides two types of remote attestation: Intel Enhanced Privacy ID (Intel EPID) attestation and Elliptic Curve Digital Signature Algorithm (ECDSA) attestation \cite{Knauth2018IntegratingRA}. Intel first introduced an EPID attestation that relied on the Intel Attestation Service (IAS), which is an Intel-specific entity that is responsible for the verification of the validity of the platform. The platform consists of an SGX application, an SGX application enclave and the Intel Quoting Enclave. The application on the user platform creates the SGX application enclave to perform some sensitive code using an encrypted area of the memory. Quoting enclave is a special enclave on every Intel SGX processor which is responsible for local and remote attestation. 

The RA flow starts with a remote user sending a request, with a nonce to guarantee freshness, to the SGX platform. The SGX application receives the request and forwards it to its application enclave, along with the Quoting Enclave's identity and the nonce. The application enclave then calls the EREPORT instruction to create a cryptographic REPORT that includes enclave-specific information such as enclave content, stack and heap, location of each page within the enclave, security flags of the enclave, etc. The enclave identity (MRENCLAVE) which is a 256-bit digest of the log and the nonce are also included in the generated report. The SGX application then forwards this report to the Intel Quoting Enclave, which receives a cryptographic key (report key) by calling the EGETKEY instruction. It then verifies the report with the report key. This report verification is defined as local attestation, since the Quoting Enclave and the application enclave are on the same platform. Once local attestation is successfully performed, the Quoting Enclave signs the report with a \textit{secret attestation key} that is provided by Intel. This signed report is called a \textit{Quote} and can then be verified using a public key on the client side. 

The attestation key and the verification protocol are different for EPID and ECDSA attestation. Initial versions of Intel SGX focused on privacy-sensitive client platforms which used EPID. EPID is an anonymity-preserving group signature scheme that is designed to prevent tracing an attestation back to the individual processor that created it. This allows systems to be identified as genuine SGX platforms without revealing their identity during verification. Instead of publishing the verification keys, Intel set up an Attestation Service (IAS) to verify quotes. When the remote attestation evidence \textit{Quote} is sent back to the remote client, she can forward it to the IAS, which replies with an attestation verification report, confirming or denying the authenticity of the quote. 
ECDSA-based attestation, on the other hand, is an alternative attestation model that was later introduced by Intel. It allows third parties to build their own non-Intel attestation infrastructure. It is useful for cloud service providers who deliver applications in a distributed fashion and cannot rely on a single point of verification. As a result, ECDSA uses 256-bit Elliptic Curve secret and public key pairs, where the secret key can be used by the Quoting Enclave to generate quotes, and public key can be distributed among cloud clients and used to verify quotes. Therefore, Intel’s participation in the attestation flow is no longer required. While Intel already provides a reference implementation for ECDSA-based attestation along with a software library to generate and verify quotes, it gives third parties the freedom to modify the attestation protocol and write their own quoting enclave. 

Intel Trust Domain Extensions (Intel TDX), which was recently introduced to help deploy hardware-isolated, virtual machines (VMs) called trust domains (TDs) also relies on an ECDSA remote attestation protocol to attest its Intel TD modules. Intel TDX's RA follows the same protocol as Intel SGX. It utilizes a TD-quoting enclave that is responsible for generating quotes. \text{Only at the TD-creation}, TD requests CPU to generate a REPORT using a new instruction SEAMREPORT, which are then signed by the Quoting Enclave \cite{TDX}.

Similarly, AMD SEV, which aims to prove the correct deployment of virtual machines to the cloud customer at the VM creation time, introduces a RA protocol. Each AMD platform contains a chip-unique signing key called the Chip Endorsement Key (CEK). CEK is an ECDSA
key, that is generated from CPU-specific secrets stored in one-time programmable fuses (OTP fuse) in the CPU \cite{Buhren2019InsecureUP}. The CEK is signed by the AMD SEV Signing Key (ASK), which is signed by the AMD root signing key (ARK). A remote client can obtain the public component of the AMD signing key for verification. The programming and management of these keys are handled by
the SEV firmware running AMD Platform Security Processor (PSP) and are stored inside the PSP that uses its own private isolated memory.

\section{ADSO Adversarial Model}\label{sec:ADSO}

As presented in the previous section, the state-of-the-art secure processors commonly rely on the usage of a \textit{single} private signing key for remote attestation, upon
which the security of the remote attestation protocol, or \textit{the most fundamental root of trust} between the cloud provider and remote client depends. The secure storage and usage of these keys are therefore crucial to maintain the reliability of the remote TEE. In order to ensure this, the current secure processor architecture technology relies on the hardware isolation principles and access control mechanisms that it fundamentally implements. While theoretically and technically intact, these hardware isolation primitives are not sufficient to prevent private data from leaking through side channels, e.g., other sources of information, that can still be observed and used by a malicious adversary to extract sensitive information. 
In fact, in recent years, it has been shown that Intel SGX is vulnerable against a wide range of side channel attacks\cite{iago2013, Lee2017HackingID, Weichbrodt2016AsyncShockES, vanbulck2017sgxstep, vanbulck2019tale, Bulck2018NemesisSM, Gyselinck2018OffLimitsAL, Huo2020BluethunderA2, Wang2017LeakyCO, Bulck2018ForeshadowET, Moghimi2017CacheZoomHS, Gtzfried2017CacheAO, Schwarz2017MalwareGE, Brasser2017SoftwareGE, Dall2018CacheQuoteER, Moghimi2018MemJamAF, Lee2017InferringFC, Evtyushkin2018BranchScopeAN, Kocher2019SpectreAE, Lipp1999Meltdown, Chen2020SgxPectreSI, Schaik2020CacheOutLD, Schaik2019RIDLRI, Schwarz2019ZombieLoadCD, Ragab2021CrossTalkSD, Jang2017SGXBombLD, Kim2014FlippingBI, Zhang2019TeleHammerAF, Murdock2020PlundervoltSF}. 

Attacks such as Spectre\cite{Kocher2019SpectreAE} show how the speculative buffer can leak private information, and the secret seal keys and attestation keys from Intel signed quoting enclaves can be extracted. Similar to Spectre, Meltdown\cite{Lipp1999Meltdown} exploits the out-of-order execution property of modern CPUs to leak private information. Cache-based timing attacks (such as ``Sneaky Page Monitoring''~\cite{Wang2017LeakyCO}, CacheZoom attack\cite{Moghimi2017CacheZoomHS}, MemJam attack\cite{Moghimi2018MemJamAF}), exploit the cache-hierarchy system and the cache access patterns in the system state which are measurable from outside the protected application by the untrusted OS. Dall \etal\cite{Dall2018CacheQuoteER} also used Prime+Probe and attacked Intel’s provisioning enclave which breaks EPID’s (Intel's algorithm used for attestation while preserving the privacy of the trusted system) unlinkability property. Similarly, Buhren \etal \cite{Buhren2019InsecureUP} showed that it is possible to extract AMD's
critical CPU-specific keys that are fundamental for the security of
its remote attestation protocol, and that possession of a private signing key is sufficient to perform the attacks, regardless of whether the key belongs to the attacked platform or not. We may conclude that the private keys that are used for attestations leak due to side channel attacks.

Under these circumstances, how can we rely on the hardware isolation assumptions for the maintenance of these private keys? In order to provide a secure remote attestation protocol, we have to consider an adversary with the strongest capabilities. Let us consider an \textbf{All Digital State Observing (ADSO)} adversary that
\begin{itemize}
    \item can compromise and alter the OS, run own enclave code, and can execute or interact with instantiations of the RA enclave,
    \item can observe all digital state, which includes all intermediate digital values computed by the RA enclave as well as all digital storage together with register values, permanent storage, and fused (endorsement) keys,
    \item cannot circumvent hardware isolation in that it must adhere to the access control implemented for the RA enclave. In particular, the ADSO adversary cannot circumvent the usual access and tamper with values computed inside the RA enclave or stored in the on-chip digital storage.

\end{itemize}

Notice that the strong adversary is not physically present and breaking the processor or adding a hardware Trojan. We assume that the secure processor keeps on functioning according to its specification. In order words, we rely on only verifiable computation based on access control implemented in hardware, but not confidential computing provided by the processor. These capabilities give the adversary the opportunity to steal digital keys and perform impersonation attacks. We argue that we can only achieve secure remote attestation if we can design a protocol that is intact in the presence of an ADSO adversary.

\section{Forward Security and Key-Insulated Schemes}~\label{sec:crypto_related_work}

In an ADSO adversarial model, key leakage is inevitable, so we have to break the lifetime of the signature scheme into multiple sessions and update the secret key at the beginning of every session. This allows for each session to have its own session key. Forward-secure signature schemes aim to protect the security of past session keys even when the current session key is leaked. A good survey on forward security can be found in~\cite{Itkis2004}. On the other hand, these forward-secure schemes have a significant shortcoming when it comes to preventing the exposure of future keys, as all the future keys can be derived from and, therefore, leaked using the current session key. 

In order to prevent the leakage of future as well as past session keys, key-insulated schemes (KIS)~\cite{DodisKatzXuEtAl2002,DodisKatzXuEtAl2003,
HanaokaHanaokaShikataEtAl2005,Franklin2006,WatanabeShikata2016} have been introduced. These schemes typically have an architectural design with a \textit{user} and a \textit{base}. The secret keys are held in shares by the user and its base, and all secret session keys correspond to one universal public key. Hence, the public key does not need to be updated frequently, while the secret session keys are refreshed constantly. A $(t,N)$-KIS ensures the security of the remaining $N-t$ session keys when at most $t$ out of $N$ session keys are exposed~\cite{DodisKatzXuEtAl2002}. When $t=N-1$, we call it a \textit{perfectly} key-insulated scheme. Also, no signature forgery of any session is possible when an attacker only compromises the base in a key-insulated scheme because session keys are stored in the user and the base using secret sharing techniques.

Applying the existing key-insulated signature schemes to our system, the processor and the secure co-processor can be viewed as the user and its base in the KIS, respectively. However, in our ADSO adversarial model, where attackers can observe all states in the processor, the attackers can effectively \textit{steal every observed session key} and use it to \textit{forge a valid signature of the compromised sessions}. Hence, key-insulated signature schemes fail to deliver the security requirements we want for \textit{every single session (either compromised or uncompromised)}. We may have to combine KIS with a synchronization mechanism between the signer and the verifier to enforce the expiration of a session key to limit the damage of a compromised session to only a short period. 

In our proposed RA scheme, we introduce One-time Signature with Secret Key Exposure (OTS-SKE), in which we strengthen the security properties of key-insulated schemes. In an OTS-SKE scheme, an attacker \textit{cannot} forge a valid signature of a \textit{new} message for \textit{any} sessions even if \textit{all} session keys are leaked, while key-insulated schemes provide no security for the compromised sessions. Moreover, in our adversarial model, we assume no attacker is present on the secure co-processor, so no secret sharing between the co-processor and the processor is needed. 

\begin{figure}[!t]
\centering 
\includegraphics[width=\columnwidth]{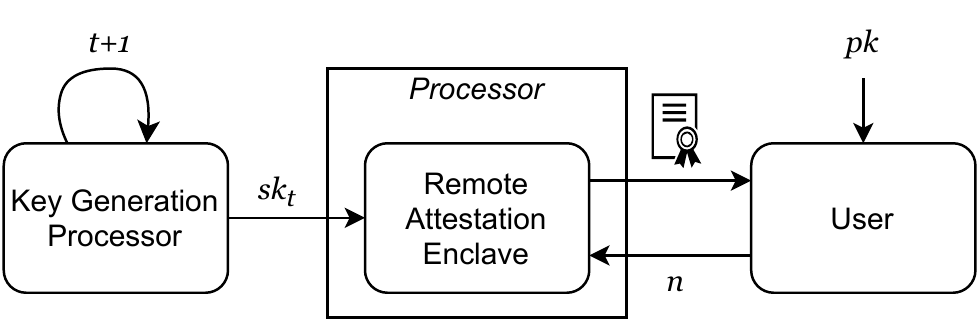} 
\caption{Overview of the proposed RA scheme. A remote user makes an attestation request sending a random nonce $n$ to the RA enclave on the processor. We have a secure co-processor that is specifically dedicated to the generation of session keys. The secret key that is used by the RA enclave is updated after each signing session by the co-processor. The RA enclave creates a RA quote with the session key $sk_t$ and user's nonce $n$ and the remote user uses the public key to verify it.}
\label{fig:intro_fig}
\end{figure}
Figure \ref{fig:intro_fig} gives a high-level overview of how such a scheme can be implemented in a secure processor architecture remote attestation context. We have a user who makes an attestation request to the RA enclave on the processor. The remote user uses the public key of the RA enclave to verify the signatures, which does not change over time. The secret key that is used by the RA enclave is updated after each signing session, and these secret keys are generated and provided by a co-processor that is separated from the RA enclave. We stress that this key generation entity has to be secure and completely isolated from the rest of the system, to protect the past and future keys. 

\section{Proposed Remote Attestation Protocol for Secure Processors}\label{sec:main_scheme}

Our purpose is to design a remote attestation protocol that is secure in the presence of an ADSO adversary. In order to do this, we need a new primitive that uses one-time signatures with forward and backward security. Given this, the next sections describe the design requirements, implementation of this new signature scheme, and how it is contextualized in state-of-the-art secure processor architectures.


\subsection{Design Requirements}

Based on our adversarial setting, we have the following requirement: we have to prevent the leakage of (1) future and (2) past secret session keys:

\subsubsection{Protection of Future Session Keys} In order to satisfy the first requirement, to prevent the leakage of future session keys, we notice that the generation of the secret keys must be done in an isolated environment. This can be done in two possible ways. The first option is to generate a number of private keys offline and store them in memory in an encrypted form. In runtime, at the time of signing, the secret session key can be retrieved from the memory. This option can be problematic in a way that a finite number of secret keys needs to be generated at the initialization phase. Once the processor runs out of session keys, a new sequence of secret keys along with a new public key need to be generated in a secure offline setting again. Secondly, the storage of these keys has to be designed carefully, in such a way that the retrieval of a session key does not leak the future session keys. While this option might be considered, another option to achieve isolated key generation is to have a physically isolated co-processor, that is separated from the main processor (where signing entity is implemented) and all other entities on the chip. We can leverage this physically isolated processor to generate secret signing keys in runtime simultaneously, which will then be used by the signing entity, i.e. the remote attestation enclave. In our RA protocol, we will use this type of isolation for key generation. This implies that, as a hardware root of trust, we need a separate isolated secure processor, that is \textit{only} responsible for the generation of session keys. The signing enclave, in fact, is not needed to be modified and can keep on functioning in its default setting. On the other hand, the specifications of this secure key generation co-processor must be carefully defined to prove isolation from a potentially insecure processor. We require that there must be only one directional information flow from the key generation processor to the outside world. This is important because we have to minimize the attack surface of this co-processor which essentially contains sensitive data to be used for key generation, by removing any user-level communication channels that can possibly enter the chip and only keeping code and data related to key generation. One may argue that the signing module can also be offloaded to an isolated processor. As stated before, we want to contain as little secure computation as possible inside the isolated processor. We stress that no other entity from the user platform, or any entity that is in communication with the user enclave should be contained in the secure isolated processor. The only entity that has to be kept private is the key generation module. The privacy of the remote attestation enclave is not a concern for us. We only make the assumption that the computation of remote attestation enclave in the processor cannot be tampered with, but it can be observed by an ADSO adversary and is not private. The leakage of the digital state of the remote attestation enclave essentially does not impact the security of our RA protocol because we renew the secret signing keys as soon as they are used in the signing enclave. 

\vspace{2mm}
\subsubsection{Protection of Past Session Keys} The second requirement we have, protecting the past keys, needs to be fulfilled in order to prevent impersonation attacks that can be performed by an ADSO adversary who is able to observe session keys once they enter the insecure processor. Once a signing key is used by the remote attestation enclave, we assume that it is automatically leaked to the ADSO adversary who can later use it. Under this assumption, how can we prevent the adversary from impersonating the cloud provider since the remote user cannot know whether the corresponding session key (and signature) is fresh? Hence, we need the signing key to be used indeed only \textit{once} and the remote user should be able to verify that it is fresh and used for the first time. We guarantee freshness by generating a session key that is unique to the remote user based on a random nonce that she sends. We do this by generating a sequence of secret keys for each session in an isolated environment, and based on the random nonce received, we select a subset of keys from this sequence. We combine this selected subset and create a single unique key, which is used by the remote attestation enclave for signing. The ADSO adversary can observe this subset of keys, but despite leaking them, he cannot forge a signature since the random nonce selected by a remote user is different for each session and a forged signature will fail during the verification if it does not match with the user's nonce. The adversary cannot create another key unique to a different nonce, because he did not observe the rest of the secret keys and only knows the subset of keys related to an older nonce. However, when we introduced the secure key generation co-processor, we stated that the information flow must be \textit{one-directional}. Therefore, how can the secure co-processor output a unique subset of keys if we do not allow any input (e.g. random nonce) to it? We do this by introducing an additional module to our hardware root of trust along with the secure key generation processor: a special oblivious transfer memory between the secure co-processor and the RA enclave as a buffer, which is implemented by its own access interface. We can allow the secure key generation co-processor to generate the whole sequence of keys for each session, and store them in this memory periodically. Whenever the RA enclave receives a request from a remote user with a random nonce, it will want to access this memory to read a certain subset of keys related to this nonce. We can design this oblivious transfer memory with a one-time read property: once a certain subset of keys is read, the rest of the keys that are not selected are consequently erased, not by reading them but by overwriting them. Even if the ADSO adversary is able to access this memory, he sees no digital value as a result of this erasure. Again, we stress that we restrict the ADSO adversary to a software-ADSO, hence, it must use the interface to get access to the memory (and cannot use e.g. a heat map to observe another subset of keys once one subset is already read). Therefore, we have a secure co-processor that generates keys and stores these into the oblivious transfer memory that it maintains. If the key generation is observed by the software-ADSO then he can perform impersonation attacks, but if it is not observed, then this special oblivious transfer memory hides the preprocessed keys from the adversary by having this type of access interface.

\subsection{Definition of OTS Scheme with Secret Key Exposure}

A basic building block for designing our remote attestation protocol is a new primitive, called a Public Key Session based One-Time Signature (OTS) Scheme with Secret Key Exposure, denoted OTS-SKE scheme for short. The idea is to have (1) one (universal) public key that can be used to verify all session signatures, (2) each session generates at most one signature with its own secret session key that is uniquely generated based on a random nonce sent by the remote user, and (3) this unique session key is exposed to the adversary for free. 

\vspace{2mm}

\noindent \subsubsection{Implementation} 
An OTS-SKE scheme ${\cal S}$ consists of three procedures
$${\cal
S}=(\Call{KeyGen}{},  \Call{Sign}{}, \Call{Verify}{}):$$ 

\noindent
{\bf Key generation.} Based on a security parameters $\lambda$, \Call{KeyGen}{} generates a public key $pk$ together with session secret keys
$$sk_i = \{sk_{i,j}\}_{j=0}^{q-1}$$
and auxilairy variables $aux_i$
for each session $i\in \{0,\ldots, N-1\}$ and a-priori fixed parameter $q$. We have 
$$
(pk, \{sk_i, aux_i\}_{i=0}^{N-1}) 
\leftarrow \Call{KeyGen}{\lambda}.
$$


\vspace{2mm}

\noindent
{\bf Sign.}
\Call{Sign}{} 
takes as input the session id $i$ with session secret key $sk_i$ and auxiliary variable $aux_i$. Together with a message $M\in \{0,1\}^n$ as input \Call{Sign}{}  produces a signature $\sigma$,
$$ \sigma \leftarrow \Call{Sign}{sk_i,aux_i;M}.$$

The computation of \Call{Sign}{} is split in three steps:
\begin{enumerate}
    \item We have a keyed pseudo random permutation $\Call{PRP}{key;x}$ which, for each $key$, is a bijective mapping from strings $x\in \{0,1\}^n$ to $\{0,1\}^n$.
    We also have an injective mapping
    $\phi$ from $\{0,1\}^n$ to subsets of $\{0,\ldots, q-1\}$ (here, $q\geq n$).  \Call{Sign}{}  first selects a random $key$ and computes the subset
    $$ I= \phi(\Call{PRP}{key;M}) \subseteq \{0,\ldots, q-1\}.$$
    \item \Call{Sign}{} extracts a corresponding subset of the $i$-th session secret key:
$$sk_{i,I} = \{sk_{i,j}\}_{j\in I}.$$

\item \Call{Sign}{} uses $sk_{i,I}$ together with $aux_i$ and input message $M$ to produce a signature $\sigma'$. In order to make the dependence on the subset of the session key explicit, we write 
$$\sigma' \leftarrow \Call{Sign'}{sk_{i,I}, aux_i;M}.$$
\Call{Sign}{} returns $\sigma=(\sigma',key)$.
\end{enumerate}

\vspace{2mm}

\noindent
{\bf Verify.}
\Call{Verify}{}  outputs
$$ \{{\bf true}, {\bf false}\} \leftarrow \Call{Verify}{pk,i;\sigma,M}
$$
for a signed message $(\sigma,M)$ for session id $i$. 
Notice that the same public key $pk$ is used for all sessions.

\vspace{2mm}

\subsubsection{Correctness and Security Proofs} 
We show the correctness of the OTS-SKE scheme and prove that it is secure even if the adversary has the knowledge of subsets of session keys.

\vspace{2mm}

\noindent{\bf Correctness.} OTS-SKE scheme ${\cal S}$ is correct if for all $\sigma \leftarrow \Call{Sign}{sk_{i},aux_i;M}$ we have 
${\bf true} \leftarrow \Call{Verify}{pk,i;\sigma,M}$.

\vspace{2mm}

\noindent
{\bf Security.} Even if an adversary has knowledge of subsets of session keys 
$$ \{sk_{i,I_{i}} \}_{i=0}^{N-1}$$
together with auxiliary information $\{aux_i\}_{i=0}^{N-1}$,
the adversary cannot impersonate a signature for some session with id $i^*$ for a new message that has not yet been signed in session $i^*$. 
This security notion is 
formalized by 
GameOTS-SKE for ${\cal S}$ as the following security game:

\begin{itemize}

\item \textbf{Setup}: The challenger runs \Call{KeyGen}{} which returns
$$(pk, \{\{sk_{i,j}\}_{j=0}^{q-1}, aux_i\}_{i=0}^{N-1}).$$
The challenger gives $pk$ as well as $\{aux_i\}_{i=0}^{N-1}$ to the adversary.

\item \textbf{Query}: The adversary adaptively issues 
a sequence of 
messages $M_{i}$
at most 
one message for each session id $i$. The challenger computes 
$$I_{i} = \phi(\Call{PRP}{key_{i};M_{i}}) \mbox{ and } sk_{i,I_{i}} = \{sk_{i,j}\}_{j\in I_{i}}$$
for random $key_{i}$.

The challenger gives the extracted information $sk_{i,I_{i}}$ 
 with $key_{i}$ to the adversary (as soon as $M_{i}$ is received). 

Notice that the adversary can use this information to sign message $M_{i}$ for session $i$ by applying \Call{Sign'}{}. This may lead to multiple signatures for $M_{i}$ (since fresh randomness can be used for each signature generation). However, no signatures for other messages
$\neq M_i$

for session id $i$ can be forged if the following Guess does not succeed.

\item \textbf{Guess}: 
 The adversary selects a session number $i^* \in \{0,\ldots, N-1\}$ which refers to the session for which the adversary will want to forge a signature: The adversary outputs a signed message $(\sigma,M^*)$ for session $i^*$ such that 
 $M^*\neq M_{i^*}$

The adversary wins the game if the signature verifies, that is, 
$$
{\bf true} \leftarrow \Call{Verify}{pk,i^*;\sigma,M^*}.
$$

\end{itemize}
In this game, $\mathcal{A}$ is called an OTS-SKE-EUF-CMA (OTS-SKE 
Existential UnForgeability

under Chosen Message Attack) adversary.

If $\mathcal{A}$ wins GameOTS-SKE with probability $\geq \epsilon$ in time $\leq T$, then we call $\mathcal{A}$ a $(T,Q_H,Q_P,\epsilon)$-OTS-SKE adversary for ${\cal S}$, where $Q_H$ and $Q_P$ are the maximum number of queries allowed to be made by $\mathcal{A}$ to a hash function oracle and PRP oracle in GameOTS-SKE. We say scheme ${\cal S}$ is $(T,Q_H,Q_P,\epsilon)$-secure against OTS-SKE-EUF-CMA attacks if no $(T,Q_H,Q_P,\epsilon)$-OTS-SKE adversary exists.

\subsection{Proposed Remote Attestation Protocol}

Our goal is to offer secure remote attestation even when an ADSO-adversary is present. We want to show that an application enclave AppEnc can have its computed result signed for a remote user by a remote attestation enclave, called RAEnc for short. We require that the ADSO adversary, who can (1) observe all digital (intermediate) state of the AppEnc and RAEnc computations (including all digital storage) and who can (2) ask the RAEnc to sign whatever it wants, cannot forge a signature on behalf of AppEnc for the remote user. That is, even with access to all digital state, the ADSO adversary cannot forge a signature for a remote user of a malicious result $R$ (for which AppEnc had not asked for it to be signed) such that it passes the signature verification by the remote user.

\begin{figure}[!t]
\centering 
\includegraphics[width=\columnwidth]{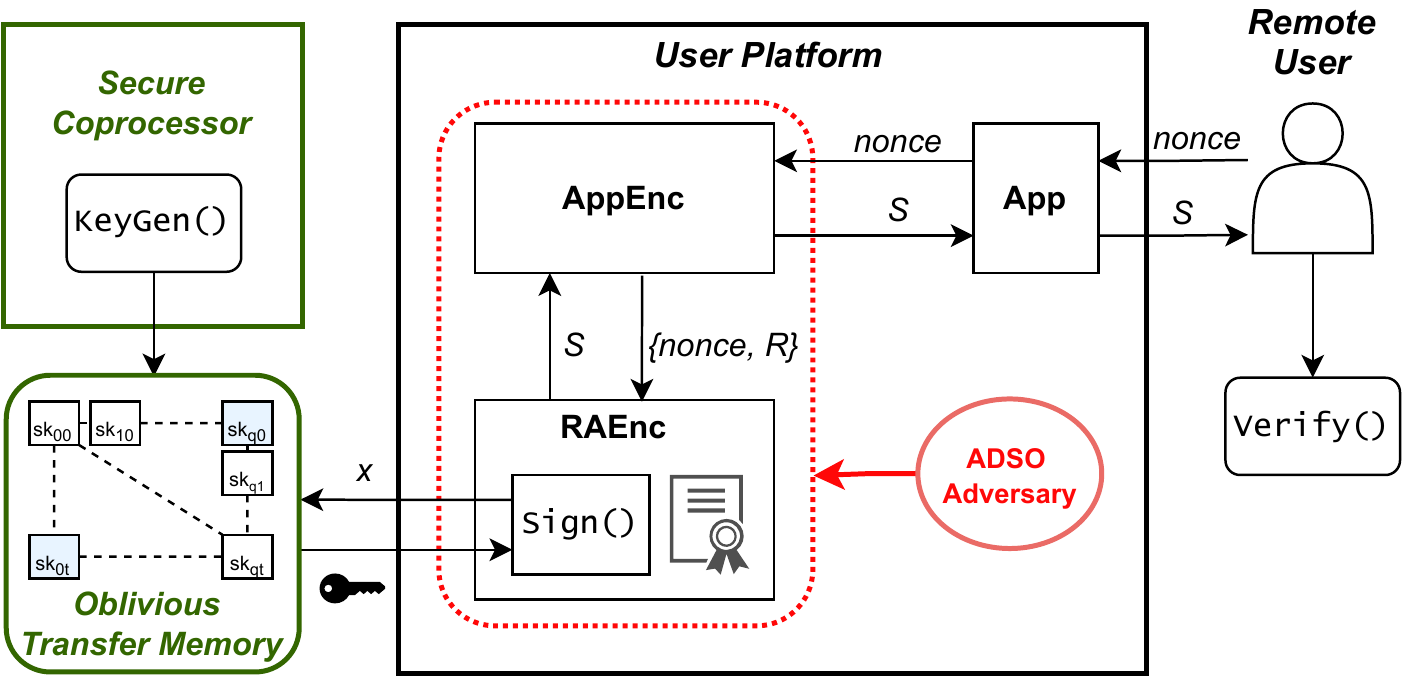} 
\caption{Proposed RA protocol. The remote user starts the RA process by sending a random nonce $nonce$. The AppEnc forwards this request along with its computed result $R$ to RAEnc. In order to retrieve a signing key, RAEnc computes a message $M$ that takes the hash of its measurement and AppEnc's measurement along with the result $R$. It then makes a read request to the Oblivious Transfer Memory by sending its input $x = \Call{Hash}{nonce, M}$. The Oblivious Transfer Memory stores the complete set of keys for a session, generated by the secure key generation co-processor. As soon as it receives the read request from RAEnc, it extracts a unique subset of keys related to the input $x$ and the measurement of RAEnc, the enclave that queries the Oblivious Transfer Memory, and immediately erases the keys that were not included in the subset. RAEnc uses this key to create a signature and sends it back to the remote user, who can then verify the signature using the public key.}
\label{fig:RA_protocol}
\end{figure}

Figure \ref{fig:RA_protocol} depicts our solution. We have 3 main functions of the OTS-SKE scheme: \Call{KeyGen}{}, \Call{Sign}{} and \Call{Verify}{}. The secure isolated processor implements \Call{KeyGen}{} to generate a sequence of session keys. The \Call{Sign}{} functionality is implemented by the RAEnc, and the remote user uses \Call{Verify}{} to verify the result sent by the RAEnc. 
Algorithm \ref{alg:RA} describes the signature generation process for the RA scheme. The signature generation contains 3 main entities: \Call{KeyGenProcessor}{}, \Call{EOTMem}{.} and \Call{RAEnc}{.}. \Call{KeyGenProcessor}{} is responsible for the key generation module. \Call{EOTMem}{.} is implemented as an enclave call that handles the oblivious transfer memory's read requests and implements its access control mechanism. Finally, \Call{RAEnc}{.} handles the RA requests from the remote user and is responsible for signature generation.

There are 2 working modes of our RA protocol: initialization and runtime. During the initialization, the secure key generation processor uses \Call{KeyGen}{} to generate a public key $pk$ and initializes the default state of the session counter $ctr$ to 0 (Line 3--5). RAEnc, on the other hand, receives this $pk$ from the key generation processor and binds it together with its measurement $\mbox{MR}_{RAEnc}$ in a certificate (Line 2--4). The $pk$ is known by the remote user and used during attestation verification. After the initialization, the essential components of the RA protocol are set up. 

\begin{algorithm}
\caption{Remote Attestation}\label{alg:RA}

We assume a OTP-SKE-EUF-CMA secure bilinear-map based OTS-SKE scheme ${\cal S}=(\Call{KeyGen}{}, \Call{Sign}{}, \Call{Verify}{})$ where \Call{Sign}{} is defined by procedure $\Call{Sign'}{}$. $\Call{KeyGen}{}$ is implemented in a secure isolated processor and $\Call{Sign}{}$ is performed on the RAEnclave. Signing is implemented using our OTS-SKE scheme which only uses a subset of the session keys for each session and the subset choice is $nonce$ dependent.
We store the $ctr$ variable in the persistent on-chip store of the isolated key generation processor, which represents the session counter.

\begin{algorithmic}[1]
\Procedure{KeyGenProcessor}{}
\If{$Mode ==$ initialization}
\State $pk \leftarrow \Call{KeyGen}{\lambda}$
\State Set  $ctr=0$
\State \Return $pk$
\ElsIf{$Mode ==$ runtime}
\State $ctr$ += 1
\State $i=ctr$
\State $key = (\{sk_{i,j,b}\}_{j=0,b=0}^{n-1,t-1}, aux_{i})\leftarrow$\Call{KeyGen}{$\lambda$}
\State Send $\{aux_{i}, i\}$ to RAEnc
\State Store $sk_i=\{sk_{i,j,b}\}_{j=0,b=0}^{n-1,t-1}$ in oblivious transfer memory
\EndIf
\EndProcedure

\end{algorithmic}

\begin{algorithmic}[1]
\Procedure{EOTMem}{$x$}
\State Compute $I = \phi(\Call{Hash}{x,\mbox{MR}_{caller}})$
\State Read  $y=(Mem_j)_{j\in I}$
\State Erase  $(Mem_j)_{j\notin I}$
\State \Return  $y$
\EndProcedure

\end{algorithmic}

\begin{algorithmic}[1]
\Procedure{RAEnc}{}
\If{$Mode ==$ initialization}
\State $pk \leftarrow \Call{KeyGenProcessor}{}$
\State Produce a certificate binding $pk$ and $\mbox{MR}_{ra}$ 
\ElsIf{$Mode ==$ runtime}
\While{{\bf true}}

\State Pop a signing request from the queue  
\State \hspace{.5cm} $(\mbox{MR}_{app},R, \mbox{RemoteUser})$ 
\State $M=\Call{Hash}{\mbox{MR}_{RAEnc},\mbox{MR}_{app},R}$

\State {\bf receive} $nonce$ from RemoteUser 
\State $x=\Call{Hash}{nonce,M}$
\State Read $\{aux_{ctr}, ctr\}$ from \Call{KeyGenProcessor}{}
\State $\{key_I\} \leftarrow \Call{EOTMem}{x}$
\State  $k \leftarrow$ Combine $key_I, aux_{ctr}$
\State $\sigma' =\Call{Sign'}{k;M}$

\State $S \leftarrow (ctr,\sigma',\mbox{MR}_{RAEnc},\mbox{MR}_{app},R)$
\State \hspace{-2mm} {\bf send} 
$S $ to RemoteUser

\EndWhile
\EndIf
\EndProcedure
\end{algorithmic}

\end{algorithm}

During runtime, \Call{KeyGenProcessor}{} increments the session counter $ctr$ and generates the complete set of secret keys $sk_i$ along with the auxiliary information $aux_i$. It directly forwards $aux_i$ with the session $ctr$ to RAEnc, and it stores the set of subkeys that constitute session secret key $sk_i=\{sk_{i,j,b}\}_{j=0,b=0}^{n-1,t-1}$ in the oblivious transfer memory (Line 6--11).

$\Call{EOTMem}{.}$ is implemented as an ECall (enclave call) that handles the read requests for the oblivious transfer memory. It computes the hash of input $x$, concatenated with the caller enclave's measurement $\mbox{MR}_{caller}$ to compute the set $I=\phi(\Call{Hash}{x, \mbox{MR}_{caller}})$ to extract a subset of keys from memory that is related to $I$ (Line 2). Here, we define $$\Call{PRP}{i;M} = \Call{Hash}{\Call{Hash}{nonce_i,M}, \mbox{MR}_{RAEnc}},$$
where $nonce_i$ is the nonce received from RemoteUser for session $i$. Regarded as a function of $M$ a collision resistant hash function $\Call{Hash}{nonce_i,M}$ cannot be distinguished from a pseudo random permutation with non-negligible probability. Therefore, we may use this for our PRP and fit the definition of the OTS-SKE scheme. As soon as the subset of secret keys $y$ is extracted, the rest of the secret keys that are not included in subset $I$ are erased from the oblivious transfer memory, and $y$ is returned to the caller (Line 3--5).

Remote attestation session starts with RAEnc receiving a signing request. The RAEnc receives a result $R$ from an AppEnc, along with AppEnc's measurement MR$_{app}$, which needs to be signed for a remote user. RAEnc receives $R$ from AppEnc by means of a "local attestation" primitive. So far, we only discussed remote attestation but for the whole protocol to work, we also need secure local attestation. We also have to consider the possibility that this can be broken or impersonated if an ADSO adversary is present and has stolen the master key of the local attestation mechanism. Then, the ADSO adversary can remotely circumvent the application enclave and impersonate its identity. One way of circumventing this is to implement local attestation as a physical authentic channel between enclaves where the channel is hardware isolated in that the messages transmitted over the channel cannot be tampered with and the source/authenticity of messages cannot be modified. Such a physical authentic channel does not exist in Intel SGX where local attestation is implemented using crypto (AES and MAC) based on a secret key; we cannot use this because our ADSO-adversary can observe digital secret keys and thus impersonate Intel SGX's local attestation. Instead, we can either use Sanctum's \cite{Costan2016SanctumMH} solution where the security monitor implements a physical hardware isolated channel between enclaves without needing crypto and secret keys, or we can simply combine the AppEnc and RAEnc together in one single enclave such that a physical authenticated channel is inherently present (in the latter case RAEnc can be seen as a wrapper around AppEnc).

In this paper we implement the second solution where AppEnc and RAEnc must be glued together so that its measurement creates a unique message $M$. Therefore, when we write $\mbox{MR}_{RAEnc}$ and  $\mbox{MR}_{app}$ together, we mean one single measurement of the combined AppEnc and RAEnc enclave.
RAEnc will take the measurement of AppEnc and itself with the result $R$ to create message$$ M = \Call{Hash}{\mbox{MR}_{RAEnc}, \mbox{MR}_{app}, R}.$$ After this, the random nonce, that is received as a part of the RA request, is used to compute the input $x = \Call{Hash}{nonce, M}$ (Line 9--11). Meanwhile, RAEnc receives the current session's auxilary key along with the session counter $ctr$ from \Call{KeyGenProcessor}{}. In order to read the subset of secret keys from the oblivious transfer memory, it makes the $\Call{EOTMem}{x}$ call and it reads the secret key subset that is related to the input $x$ (which is unique to $nonce$ and the message $M$) from the oblivious transfer memory. It then combines this unique subset of secret keys, along with the auxiliary key $aux_{ctr}$ to generate the signing key $k$, which is used to sign the message $M$ (Line 12--15). Finally, RAEnc sends the session counter $ctr$ and the resulting signature $S$ consisting of $\sigma'$, measurement of itself (RAEnc + AppEnc) and the result $R$ to RemoteUser (Line 16--17).

\begin{algorithm}[t]
\caption{Request and Verification}\label{alg:RAuser}
We assume the remote user knows $pk$ (verified by means of a certificate issued by a trusted CA) of the OTP-SKE-EUF-CMA secure OTS-SKE scheme ${\cal S}=(\Call{KeyGen}{}, \Call{Sign}{}, \Call{Verify}{})$ used by the RA enclave.

\begin{algorithmic}[1]
\Procedure{RequestAndVerify}{}

\State {\bf send} random $nonce$  to RAEnc 
\State \hspace{-2mm} {\bf receive} 
$S$ from RAEnc
\State $(ctr,\sigma',\mbox{MR}_{RAEnc}, \mbox{MR}_{app}, R)  \leftarrow S$   
\State $M=\Call{Hash}{\mbox{MR}_{RAEnc},\mbox{MR}_{app},R}$
\State $I = \Call{Hash}{\Call{Hash}{nonce_i,M}, \mbox{MR}_{RAEnc}}$ 
\State $\sigma = (\sigma', I)$
\State {\bf return} $\Call{Verify}{pk,ctr;\sigma, M}$

\EndProcedure
\end{algorithmic}

\end{algorithm}

On the remote user's side (explained in Algorithm \ref{alg:RAuser}), after selecting a nonce $nonce$ and making a RA request to RAEnc with it, the user receives the created signature $ S =(ctr,\sigma',\mbox{MR}_{RAEnc}, \mbox{MR}_{app}, R)$ for the $i$-th session with $i=ctr$ (Line 2 and 3). This implies that even if our ADSO-adversary runs the RAEnc itself, it only learns digital state from sessions with id $\leq ctr$, and in fact only learns at most one subset of keys $\{sk_{i,j}\}_{j\in I}$ with $aux_i$ per session id $i$ (because once a subset is used for signing, $ctr$ is incremented). Also, if the adversary attempts to read the Oblivious Transfer Memory, then the returned subset of keys depends on the adversarial enclave's measurement through a hash evaluation (Line 2 in \Call{EOTMem}{}). Because of the hash collision resistance, the adversary cannot use \Call{EOTMem}{} to learn a set of subkeys that fits $\mbox{MR}_{RAEnc}$. If the adversary observes the message $M$ as supplied by the remote verifier, then he learns the related subset of keys of session $ctr$ and can create other $\sigma'$ for the same message $M$ and session id $ctr$. But this does not generate a signature for a new malicious message $M^*$. A new malicious message $M^*$ corresponds to a different subset of the session key indicated by a set $I^*$. The security guarantee of the  OTS-SKE scheme shows that the adversary cannot successfully forge a signature for $M^*$.

If $\Call{Verify}{}$ in Algorithm \ref{alg:RAuser} verifies the signature $S$ (Line 4--8), then the remote user may conclude that $R$ was indeed created by AppEnc at the processor: The chain of trust shows that the signature was created by RAEnc. Its runtime mode shows that it only signs messages $M$ that are a hash of $(\mbox{MR}_{RAEnc}, \mbox{MR}_{app})$ together with $R$.
Finally, RemoteUser selected a random $nonce$, hence, the signature is fresh (and not a replay of an older signature for the same message). We conclude that under the ADSO adversary our scheme offers secure remote attestation.

\section{Implementation and Performance Evaluation }
\label{sec:performance}
In this section, we give the details of our implementation and performance analysis for the version with compressed signatures of our bilinear map based OTS-SKE scheme that is formally described in Appendix \ref{sec:bilinear}. The pseudo-codes for the implementation of our OTS-SKE scheme's key generation, signing and verification phases are given in Appendix \ref{sec:pseudocodes}.

\subsection{Experimental Setup} 

All experiments are conducted on a PC with Intel(R) Core(TM) i7-7820HQ CPU @ 2.90GHz and 32GB memory and on Windows 10 based operating system. Intel Software Guard Extensions (SGX) is used to provide secure enclaves for the RA scheme. We constructed our bilinear map based OTS-SKE scheme using the open-source MIRACL Multiprecision Integer Cryptographic Library\footnote{https://github.com/miracl/MIRACL} that includes elliptic curve cryptography arithmetic. We used C++ language for our implementation, along with fast in-line assembly language alternatives for most performance-critical parts of our code to speed up performance, such as modular multiplication and exponentiation. We compare our key generation, signing and verification performance with our benchmark, which is the standard Elliptic Curve Digital Signature Algorithm (ECDSA) based attestation used by Intel SGX. The results in Table \ref{tab:costtable} are collected from our own experimental results. Both schemes are implemented with 256-bit security.

\subsection{Experimental Results of Bilinear Map based OTS-SKE and Comparison with Intel SGX's Remote Attestation}

\subsubsection{Runtime Cost Analysis.}
Today, the wide employment of elliptic curve cryptography (ECC) in various applications relies on a variety of implementation types from pure software or hardware implementations to hardware and software co-design. However, pure software implementations of ECC, despite offering the best ﬂexibility at the lowest cost, cannot cope with the speed demands of many application areas as general purpose processors are not designed for efﬁcient handling of ECC’s underlying ﬁnite ﬁeld arithmetic. This makes software-based implementations impractical for applications in time-constrained environments that require high throughput and processing. Considering these limitations, hardware-based implementations turn out to be the more suitable alternatives. Despite this, in this paper, we evaluate a software-based implementation of the ECC-based signature scheme that has its own computational disadvantages, but can also be potentially accelerated using HW techniques which will be discussed later.

\begin{table}[]
\caption{Runtime Cost Analysis per Remote Attestation (RA) session of the bilinear map based OTS-SKE scheme in Comparison with the ECDSA used in Intel SGX (Average over 100 runs). In the bilinear map based scheme we use $t=4$ (such that $(t/\log_2 t)\cdot 64=128$ and $128/\log_2 t = 64$). The classical security of both schemes is 256 bits and a message signed during a RA session has 128 bits. }
\label{tab:costtable}
\begin{tabular}{lr|lr}
\hline
\multicolumn{2}{c|}{\textbf{OTS-SKE}}                                            & \multicolumn{2}{c}{\textbf{ECDSA}}                                                \\ \hline
\multicolumn{1}{c}{\textbf{Operation}} & \multicolumn{1}{c|}{\textbf{Time (ms)}} & \multicolumn{1}{c}{\textbf{Operation}}   & \multicolumn{1}{c}{\textbf{Time (ms)}} \\ \hline
\textit{($v_{ij}$ Generation)}                & \textit{131.4}                          & \multirow{4}{*}{\textbf{Key Generation}} & \multirow{4}{*}{\textbf{21.2}}         \\
\textit{($aux_i$ Generation)}              & \textit{4}                              &                                          &                                        \\
\textit{($sk_{ijb}$ Generation)}               & \textit{253.2}                          &                                          &                                        \\
\textbf{Key Generation (Total)}        & \textit{\textbf{388.6}}                 &                                          &                                        \\ \hline
\textbf{Signing}                       & \textbf{3.4}                            & \textbf{Signing}                         & \textbf{23.1}                          \\ \hline
\textbf{Verification}                  & \textbf{127.3}                          & \textbf{Verification}                    & \textbf{74.2}                          \\ \hline
\end{tabular}
\end{table}

\vspace{3mm}

\noindent\textbf{Key Generation.} During key generation, in the initialization phase 
the public key 
$pk=(p,\mathbb{G},\mathbb{G}_1,e,g,g_1,g_2,h)$ is generated. During runtime, $tn$ secret keys are generated per session. For our evaluation, we experiment with 1 session and choose $N=1$  with
64-bit nonce represented by $n=32$ $t$-ary symbols, where $t=4$ ($64=n\log_2 t$). This minimizes key generation and signing run time. Key generation produces $nt = 128$ subkeys (of the session key) in total.

For $pk$, the points $g, g_1, g_2$ are randomly generated from $\mathbb{G}$ (on the elliptic curve), which takes approximately $100$ ms for each. In addition, the points are preprocessed after they are generated; this prepares them for faster computation for pairing (as well as elliptic curve arithmetic operations) for the future, which takes about $10^{2}$ ms for each. These points are generated and preprocessed only once at the beginning of key generation. Thus, despite being very costly, the overhead of the $pk$ generation can be considered negligible while working in a sufficiently large timing window (for example, if we have $N>100$ sessions). Therefore, in the runtime setup, we ignore this $pk$ generation cost.

In order to generate secret keys during the runtime of remote attestation, random elements $\beta_{i,j}\in \mathbb{Z}_p$, $i \in \{0,\ldots, N-1\}$, $j \in \{0,\ldots, n-1\}$, are first generated. As previously stated, since random point generation on the elliptic curve is costly, we reduce the cost of generating random points $v_{i,j}\in \mathbb{G}$ by computing $v_{i,j}$ = $g^{\beta_{i,j}}$ which takes approximately $4$ ms. Compared to the method of selecting $v_{i,j}$ from $\mathbb{G}$ randomly for each $i$ and $j$ ($100$ ms), this is a significant speedup of almost 25 times. 

Table \ref{tab:costtable} shows our signature scheme's performance against the benchmark. It can be seen that the generation of secret keys takes 388.6 ms in total where the computation of $v_{i,j}$ takes 131.4 and computation of the auxiliary key $aux_i$ takes 4 ms. Therefore, generating 1 secret key $sk_{i,j,b}$ takes approximately $253/128 = 1.9$ ms. The key generation of the ECDSA, on the other hand, takes 21.2 ms. 

\vspace{2mm}
\noindent\textbf{Signing.} During this phase, $q = 32$ secret keys are fetched from the secure isolated KeyGen memory based on the input nonce $I$ which includes the user's random nonce $nonce$ along with the enclave measurement. The extracted keys are sent to the remote attestation enclave and combined there which outputs a unique key to the user's nonce as well as to the measurements of RA and application enclave. By using the optimized signing in Appendix \ref{sec:bilinear}, we can use simply this unique key as the signature and avoid additional elliptic curve operations for the signing process which are expensive (e.g. pairing on the elliptic curve). As it can be seen from Table \ref{tab:costtable}, this process takes about 3.42 ms while the ECDSA takes 23.1 ms for signing. This is a significant improvement on the signing side, which can potentially become the main performance bottleneck in the runtime.

\vspace{2mm}
\noindent\textbf{Verification.} Verification contains fewer components compared to the signing phase and is performed outside the RA enclave, by the remote user. 
Despite containing fewer operations, verification is still costly ($127.3$ ms) since pairing on the elliptic curve is performed 3 times in order to verify the signature.

\vspace{2mm}

\subsubsection{Evaluation and Discussion} We have presented our experimental results for the software implementation of our bilinear-map based OTS-SKE scheme and compared its runtime overheads against the ECDSA used by Intel SGX for remote attestation. In this section, we will evaluate the benefits and performance bottlenecks of our protocol and its potential for improvements.

\noindent \textbf{Improvements and Performance Bottlenecks.} Table \ref{tab:costtable} shows that the key generation phase is the main bottleneck in our signature scheme. The main contribution we have is the use of one-time-signatures for the RA in the secure processor architectures to protect the digital secrets against an ADSO adversary. This requires us to renew the secret keys after each signing session, and in order to protect against impersonation attacks by an ADSO adversary, we generate $nt$ keys per session. This extra level of security increases the overhead of the key generation protocol. On the other hand, we require  a  secure isolated coprocessor that is responsible for the key generation module which works in parallel with the RA enclave and outputs and stores a sequence of session keys continuously. As long as the RA request period remains above $388.6$ ms, we can argue that the performance of the key generation is sufficient. Our signing cost ($3.4$ ms) is much smaller than the baseline, while bringing better security benefits. Considering the fact that the remote attestation is done only once at the enclave/VM/container creation time in state-of-the-art secure processors such as Intel SGX, AMD SEV, Intel TDX, the signing cost is relatively small. For example, Intel's EPID attestation takes $31.7$ ms for quote generation and signing, at the enclave creation which takes $24.5$ ms itself on an enclave with 5 MB of memory \cite{10.1145/3007788.3007793}. As another example, Kata container takes approx. $2.6$ seconds to launch with SEV \cite{9178442}. This shows that enclave creation itself already comes at a high cost.

\vspace{2mm}
\noindent \textbf{Potential for Acceleration.} We have shown that the elliptic curve related computations are the main overhead, that are present in the key generation and verification phases. HW-based ECC acceleration methods have been frequently explored and many hardware architectures have been proposed for faster and more compact solutions\cite{4397176,1190586,Agnew1993AnIO,Sutikno1998DesignAI,Leung2000FPGAIO,Gao1999EllipticCS, 5669264}, including several methods that exploit the computing power of graphics processing units (GPU) for ECC computations in the last decade \cite{Zhang2012AccelerationOC,Cui2014HighSpeedEC, 7555336, Pu2013EAGLAE, Szerwinski2008ExploitingTP}. For example, Pan \etal\cite{7555336} has shown that, with GPU utilization, the elliptic curve digital signature algorithm (ECDSA) can achieve $8.71 \times× 10^6$ Operations Per Second (OPS) for signature generation and $9.29 \times ×10^5$ OPS for verification. Zhang \etal\cite{Zhang2012AccelerationOC} has proposed an ECC GPU-based library for bilinear pairing called "EAGL" which can achieve 3350.9 pairings/sec on GPU at the 128-bit security level. Zhang \etal\cite{Zhang2012AccelerationOC} achieved 8.7 ms per pairing on a 1024-bit security level with GPU utilization, which is a 20$\times$ speedup compared to CPU implementations. This makes our bilinear map based OTS-SKE scheme more suitable for further acceleration  and improvements (e.g., with the use of a cryptographic processor).

\section{Conclusion}\label{sec:conclusion}

We demonstrated for the first time how to design a Remote Attestation (RA) protocol that resists an All Digital State Observing (ADSO) adversary during the signing procedure. Even with secure processor technology that implements access control using hardware isolation but without any privacy guarantees (due to the recent avalanche of attacks), our remote attestation is secure and can be used to verify computation by remote users. The new RA scheme offers a first crucial level of trust for current attacked secure processor technology.

\appendices

\section{Bilinear-Map based OTS-SKE}
\label{sec:bilinear}
\noindent In this appendix, we introduce the bilinear-map based OTS-SKE that we use. 

\subsection{Implementation} 

\noindent We begin with introducing the following definitions:\\

\noindent {\bf Bilinear map.}
Let $\mathbb{G}$ be a bilinear group of prime order $p$ and $g$ be a generator of $\mathbb{G}$. Here, size $p$ of $\mathbb{G}$ is determined (by some functional relation) by the security parameter $\lambda$ of the to-be-explained constructions. Let $e:\mathbb{G} \times \mathbb{G} \rightarrow \mathbb{G}_1$ be a bilinear map, i.e., we have the following properties
\begin{itemize}
    \item {\em Bilinear} \footnote{The definition implies $e(x,yz)=e(x,y)e(x,z)$ and $e(xz,y)=e(x,y)e(z,y)$ for all $x, y, z\in \mathbb{G}$. Also, $e(g^a,g^b)=e(g,g)^{ab}=e(g^b,g^a),$ hence, $e(x,y)=e(y,x)$.}: For all $x, y\in \mathbb{G}$ and all $a,b \in \mathbb{Z}$, $$e(x^a,y^b)=e(x,y)^{ab}.$$
   
    \item {\em Non-degenerate} \footnote{The definition implies for $x=g^a\in \mathbb{G}$, if $e(g,x)=1$, then $1=e(g,x)=e(g,g^a)=e(g,g)^a$, hence, $a=0$ (since $e(g,g)\neq 1$), i.e., $x=1$. Therefore, if $e(g,x)=e(g,y)$, then $e(g,x/y)=1$ which implies $x/y=1$ or equivalently $x=y$. The property "$e(g,x)=e(g,y)$ implies $x=y$" will be used in the security proof.}: 
    $e(g,g)\neq 1$.
   
\end{itemize}
For practical usage the bilinear map should  be efficiently computable.
The above properties can be realized by the modified Weil pairing based on supersingular curves.

\vspace{3mm}

\noindent
{\bf Hash function.}
In addition to the bilinear map $e:\mathbb{G} \times \mathbb{G} \rightarrow \mathbb{G}_1$ we use a cryptographic hash function $H: \{0,1\}^{\lceil \log_2 N \rceil + n} \rightarrow \mathbb{Z}_p$. 

$H(\cdot)$ is used in the signing algorithm
to hash an index message pair into an element in $\mathbb{Z}_p$.

\vspace{3mm}

\noindent
{\bf OTS-SKE scheme.}
Below we describe our OTS-SKE scheme
$${\cal S}=(\Call{KeyGen}{},  \Call{Sign}{}, \Call{Verify}{}):$$


\begin{enumerate}
    \item {\bf Key generation.} 
We use parameters $q=tn$ and represent index $tj+b\in \{0,\ldots, q-1\}$ as the pair $(j,b)$.
\Call{KeyGen}{} sets parameters, computes the public key, and all secret keys 
$$ (pk, \{\{sk_{i,j,b}\}_{j=0,b=0}^{n-1,t-1}, aux_i\}_{i=0}^{N-1}) \leftarrow \Call{KeyGen}{\lambda}$$ 
as follows:
\begin{itemize}
\item $(p,\mathbb{G},\mathbb{G}_1,e,g,g_2)\leftarrow IG(1^\lambda)$ where $\lambda$ is the security parameter and  algorithm $IG$ generates a suitable mathematical structure for our signature scheme. $g$ and $g_2$ are generators of $\mathbb{G}$ and $\mathbb{G}_1$, respectively. 

\item Randomly generate $\alpha \in \mathbb{Z}^{*}_p$ and set $g_1=g^{\alpha}$. Define $F(i)=g^i_1 h$ where $h$ is a random number chosen from $\mathbb{G}$. Note that $F: \mathbb{Z}_p \rightarrow \mathbb{G}$.

\item Generate $N$ secret keys $\{sk_i\}_{i=0}^{N-1}$ with auxiliary information $\{aux_i\}_{i=0}^{N-1}$ as follows: $$sk_i = \{sk_{i,j,b}\}_{j=0,b=0}^{n-1,t-1}$$
with
\begin{equation} sk_{i,j,b}=g^{\alpha}_2 F(it^n+bnt^j)^{r_i}v_{i,j} \mbox{ and } aux_i= g^{r_i}, \label{eq:sk}
\end{equation}
where  $r_i$ is a random number chosen from  $\mathbb{Z}_p$ and 
$v_{i,j} = g^{\beta_{i,j}}$, 
where $\beta_{i,j}$ are  random numbers from $\mathbb{Z}_p$  such that $\sum_{j=0}^{n-1} \beta_{i,j}=0$, or equivalently,
$\prod_{j=0}^{n-1} v_{i,j} = 1$.

\item Parameters $pk=(p,\mathbb{G},\mathbb{G}_1,e,g,g_1,g_2,h)$ are made public and  secret keys $\{sk_i\}_{i=0}^{N-1}$ are kept private. The auxiliary information $\{aux_i\}_{i=0}^{N-1}$ is kept at the signer but is not kept secret (it can be accessed by anyone who wants to). Random numbers $\{r_i\}$ and $\{v_{i,j}\}$ are deleted.
\end{itemize}

\noindent

We notice that KeyGen can be implemented in \Call{KeyGenProcessor}{} using an update rule as in~\cite{ChowHuiYiuEtAl2004} to create a continuous stream of session keys.

A deviation from~\cite{ChowHuiYiuEtAl2004} is the secret sharing mechanism based on the $\{v_{i,j}\}$, which role will become clear in the security analysis and allows us to achieve resistance against OTS-SKE-EUF-CMA attacks.

\vspace{3mm}

\item {\bf Signing.} 
We compute $B=\sum_{j=0}^{n-1} b_jt^j$ with $0\leq b_j<t$ as the pseudo random permutation output
$B = \Call{PRP}{key;M}$ for a random $key$. We define  subset $\phi(B)=\{tj+b_j\}_{j=0}^{n-1}$ of $\{0,\ldots, q-1\}$ for $q=tn$. We represent its elements by the pairs $(j,b_j)$.
We produce  
$$ \sigma' \leftarrow \Call{Sign'}{\{sk_{i,j,b_j}\}_{j=0}^{n-1},aux_i;M},$$
which signs a plaintext $M \in \mathbb{G}_1$ 
as follows:
\begin{itemize}
\item Select a random $s \in \mathbb{Z}_p$.
\item Compute $x=g^{ns}_2$.
\item Compute hash value $u=H(M,x)$.

\item Compute $y=aux_{i}^{n(s+u)}=(g^{r_i})^{n(s+u)}$.
\item Compute 
\begin{eqnarray*}
sk_i &=& \prod_{j=0}^{n-1} sk_{i,j,b_j} 
= \prod_{j=0}^{n-1} 
g^{\alpha}_2 F(it^n+b_jnt^j)^{r_i}v_{i,j} \\
&=&
g^{n\alpha}_2 \left( \prod_{j=0}^{n-1} F(it^n+b_jnt^j) \right)^{r_i}  \\
&=&
g^{n\alpha}_2 (g_1^{it^n+B}h)^{nr_i} 
=
(g_2^\alpha F(it^n+B)^{r_i})^n 
\end{eqnarray*}
and
$z=sk_{i}^{s+u}=(g^\alpha_2F(it^n+B)^{r_i})^{n(s+u)}$.
\item Return signature $\sigma=(\sigma',key)$ for \begin{eqnarray*}\sigma' &=& (x,y,z) \\
&=& (g^{ns}_2,(g^{r_i})^{n(s+u)},(g^\alpha_2F(it^n+B)^{r_i})^{n(s+u)} ).
\end{eqnarray*}
\end{itemize}

\item {\bf Verification.} \Call{Verify}{$pk,i;\sigma,M$} with $\sigma=(\sigma',key)$  verifies   signature $\sigma'=(x,y,z)$ for message $M$, where $\sigma'$ is generated during the $i$-th session:
\begin{itemize}
\item Compute $u=H(M,x)$ and $B=\Call{PRP}{key;M}$.
\item The signature verifies  if and only if 
$$e(g,z)=e(g_1,g^{nu}_2 x  y^k)\times e(y,h) \mbox{ with } k= it^n+B.$$ 
\end{itemize}
\end{enumerate}

\subsection{Correctness} 

The correctness of this check follows from
$g_1= g^\alpha$, $x=g^{ns}_2$, $y = g^{r_in(s+u)}$, $z=(g^\alpha_2F(k)^{r_i})^{n(s+u)}$, and 
\begin{eqnarray*}
&& e(g^\alpha,g^{nu}_2 x  y^k)\times e(y,h) \\
&=& 
e(g^\alpha,g^{nu}_2 g^{ns}_2  (g^{r_in(s+u)})^k)\times e( g^{r_in(s+u)},h)
\\
&=&
e(g^\alpha,g^{n(u+s)}_2   (g^{r_in(s+u)})^k)\times e(g^{r_in(s+u)},h) \\
&=& (e(g^\alpha,g_2   g^{kr_i})\times e(g^{r_i},h))^{n(s+u)}.
\end{eqnarray*}
Since $e(a^c,b)=e(a,b^c)$, the right hand side is equal to 
\begin{eqnarray*}
&& (e(g,g^\alpha_2 g^{\alpha k r_i }) \times e(g,h^{r_i}))^{n(s+u)} \\
&=& (e(g,g^\alpha_2  (h g^{\alpha k})^{r_i}))^{n(s+u)}
= e(g, (g^\alpha_2  (h g_1^{ k})^{r_i})^{n(s+u)}) \\
&=& e(g, (g^\alpha_2 F(k)^{r_i})^{n(s+u)})
= e(g,z).
\end{eqnarray*}

\subsection{Security}

Our security analysis will apply the forking lemma~\cite{pointcheval2000security} in order to reduce an OTS-SKE adversary $\mathcal{A}$ to being able to solve the Computational Diffie-Hellman Problem (CDHP) in $\mathbb{G}$. We first give some background:

\vspace{3mm}

\noindent
{\bf Forking lemma~\cite{pointcheval2000security}.} Our scheme corresponds  to a generic digital signature scheme~\cite{pointcheval2000security}, i.e., given an input message $M$, the scheme 
produces a triple $(\sigma_1, u, \sigma_2)$ where $\sigma_1$ ($=x$ for our scheme) randomly takes its values in a large set (in our case 
$x\in\mathbb{G}$), $u$ is
the hash value $H(M, \sigma_1)$ 
and $\sigma_2$ ($=(y,z)$ for our scheme) only depends on $\sigma_1$,  message $M$, $u$, and secret keys (but does not depend on other randomly selected values). We notice that $u$ does not need to be transmitted as part of the signature since it can be computed by the verifier by applying the hash function.

We denote by $Q_{H}$ the number of queries that ${\cal A}$ can ask a random oracle which models the hash functionality.
Suppose that within a time bound $T$ and with probability $\epsilon\geq 7Q_{H}/2^\lambda$, where $\lambda$ is the security parameter of the scheme,
${\cal A}$ can produce a valid signature $(M,\sigma_1, u, \sigma_2)$.
Then
there exists another algorithm which has control over ${\cal A}$ and produces two valid signatures $(M,\sigma_1, u, \sigma_2)$ and $(M,\sigma_1, u', \sigma'_2)$ such that $u\neq u'$ in expected  time $T'\leq 84480 \cdot T Q_{H}/\epsilon$. The new algorithm is in control of the random oracle and simulates ${\cal A}$ for a first hash function $H(.)$ with $u=H(M,\sigma_1)$ and for a second hash function $H'(.)$ with $u'=H'(M,\sigma_1)$. 

\vspace{3mm}

\noindent
{\bf Computational Diffie-Hellman problem.}
Our security analysis shows that the two valid signatures can be algebraically combined  allowing us to solve the Computational Diffie-Hellman Problem (CDHP) in $\mathbb{G}$: Given a triple $(g,g^a,g^b)\in \mathbb{G}^3$, compute $g^{ab}\in \mathbb{G}$.\footnote{We say that the $(t',\epsilon)$-CDH assumption holds in $\mathbb{G}$ if no $t'$-time algorithm has advantage at least $\epsilon$ in solving the CDHP in $\mathbb{G}$.}

The following theorem states the reduction of OTS-SKE-EUF-CMA security to CDHP.

\vspace{3mm}

\noindent 
\begin{theorem}
Let ${\cal S}$ be the bilinear map based OTS-SKE scheme with $N$ sessions. Let $\mathcal{A}$ be a $(T,Q_H,Q_P,\epsilon)$-OTS-SKE adversary for ${\cal S}$ with $\epsilon\geq 7Q_H/2^\lambda$, where $\lambda$ is the security parameter of ${\cal S}$. Then, there exists an algorithm that solves CDHP in $\mathbb{G}$ in expected time $\leq 84480\cdot 2TQ_H(Q_P+1)N/\epsilon$.
\end{theorem}

\vspace{3mm}

The proof of our main result is as follows:

Let $\mathcal{A}$ be a $(T,Q_H,Q_P,\epsilon)$-OTS-SKE adversary with $\epsilon\geq 7Q_H/2^\lambda$.
We will show how to build an algorithm $\mathcal{B}$ which can solve CDHP in $\mathbb{G}$ based on $\mathcal{A}$. 

$\mathcal{B}$ takes on the role of challenger in GameOTS-SKE and plays against $\mathcal{A}$, and whenever $\mathcal{A}$ needs to compute a hash value then $\mathcal{B}$ plays the role of the hash oracle, that is, the random oracle which simulates  hash function $H(.)$; $\mathcal{B}$ keeps a list $L_H$ to store the answers of the random oracle, that is, if
$\mathcal{A}$ queries a hash value of some input $w$, then $\mathcal{B}$ checks list $L_H$ and if an entry $(w,u)$ for the query is found, the same answer $u$  will
be returned to $\mathcal{A}$, otherwise, $\mathcal{B}$ randomly generates a value $u$ from $\mathbb{Z}_p$ and outputs $u$ and 
appends $(w,u)$ to $L_H$. We only talk about algorithm $\mathcal{A}$ querying the hash oracle. As it turns out $\mathcal{B}$ does not query its own  hash oracle at all.

$\mathcal{B}$ is also in charge of the random tape used by $\mathcal{A}$, i.e., $\mathcal{B}$ is able to replay $\mathcal{A}$ for the same random tape and with a random oracle that simulates a different hash function $H'(.)$. 
As we will see, by exploiting this, $\mathcal{A}$ can be used by $\mathcal{B}$ to solve CDHP.

Finally, $\mathcal{B}$ also plays the role of the PRP oracle, that is, the random oracle which simulates the pseudo random permutation \Call{PRP}{}. 
Now $\mathcal{B}$ selects at the start of its execution, for each $i\in \{0,\ldots, N-1\}$, 
a set of distinct random values  $$Set_{PRP}(i)=\{B^*_0, \ldots, B^*_{Q_P+1}\} 
\subseteq \{0,1\}^n$$
together with a random $key_i$.
    Notice that the size of each set is $Q_P+2$. The set is used to answer PRP oracle queries from both $\mathcal{A}$ as well as $\mathcal{B}$ (it queries itself).
$\mathcal{B}$ maintains a list $L_P$ of triples $(i,M,B)$: 
If
$\mathcal{B}$ or $\mathcal{A}$ queries a PRP value for some input $(key_i;M)$, then $\mathcal{B}$ checks list $L_P$ and if an entry $(i,M,B)$ for the query is found, the same answer $B$  will
be returned to itself, otherwise, $\mathcal{B}$ randomly generates a value $B$ from 
$$Set_{PRP}(i)\setminus \{ B' : (i,M',B')\in L_P\}$$  and outputs $B$ and
appends $(i,M,B)$ to $L_P$. Notice that the PRP oracle only outputs values from $Set_{PRP}(i)$ for each $i$. This is possible because $\mathcal{A}$ makes at most a total of $Q_P$ PRP queries and, as we will see below, $\mathcal{B}$ makes at most $2$ PRP queries per session (one during Setup and one during Guess).

Let  
$$g_1=g^a \mbox{ and } g_2=g^b.$$ 
We define algorithm $\mathcal{B}$ with its interactions with ${\cal A}$ as follows:
\begin{itemize}
    \item \textbf{Setup}:
    $\mathcal{B}$ selects a random session id $i^*\in \{0,\ldots, N-1\}$ together with 
    one random element $B^*$ from $Set_{PRP}(i^*)$. 
    %
    $\mathcal{B}$ selects a value $\alpha' \in \mathbb{Z}_q$ at random and defines $$h=g^{-k^*}_1 g^{\alpha'} \mbox{ with } k^*=i^* t^n +B^*.$$ 
    $\mathcal{B}$ prepares 
    $pk=(p,\mathbb{G},\mathbb{G}_1,e,g,g_1,g_2,h)$
    and transmits $pk$ to $\mathcal{A}$.
    
    For each $key_i$, $\mathcal{B}$ queries the PRP oracle for a random dummy message $D_i$ with the slight adaptation that the PRP oracle does not output $B^*$ for session id $i^*$ 
    (this corresponds to the attacker choosing a new message for which a signature is constructed/guessed as we will see later).
    This results in  numbers
    $B_i$ randomly selected from $Set_{PRP}(i)$ with $B_{i^*}\neq B^*$. $\mathcal{B}$ selects random  numbers $r_i$ in $\mathbb{Z}_p$. Next $\mathcal{B}$ computes for
    $$ k_i = it^n+B_i
    $$
    the auxiliary information
    $$ aux_i = g_2^{\frac{-1}{k_i-k^*}}g^{r'_i},$$
    which is given to ${\cal A}$. Notice that $k_{i^*}\neq k^*$ because $B_{i^*}\neq B^*$. 
    Also $k_i=it^n+B_i\neq i^*t^n+B^*=k^*$ for $i\neq i^*$ because $0\leq B^*<t^n$ and $0\leq B_i<t^n$.

    \item \textbf{Query}: When $\mathcal{A}$ issues a message $M_i$ for session id $i$, then $\mathcal{B}$ replaces the triple $(i,D_i, B_i)$ with $(i,M_i,B_i)$ in the list $L_P$ of  the PRP oracle. This is allowed since $\mathcal{B}$ has used the PRP oracle only a single time for $key_i$ and, since it is chosen at random and therefore unknown to ${\cal A}$, no extra queries  have been issued to the PRP oracle for $key_i$. From the perspective of $\mathcal{A}$ it is as if the PRP oracle produced $B_i$ for $(key_i;M_i)$ (without loss of generality the dummy message $D_i$ took the shape of $M_i$).
    
    Value $B_i=\sum_{j=0}^{n-1} b_{i,j} t^j$ 
    is used to compute, as before, 
    $k_i=it^n+B_i$. Together with the random $r'_i$ previously chosen in Setup, $\mathcal{B}$
    constructs
    \begin{eqnarray*} 
    sk_{i,j,b_{i,j}} &=& g_2^{\frac{-\alpha'}{k_i-k^*}} F(it^n+b_{i,j} n t^j)^{r'_i} v'_{i,j},
    \end{eqnarray*}
    for $j\in \{0,\ldots, n-1\}$, 
    with
    $$
    v'_{i,j}=g^{\beta'_{i,j}} \mbox{ such that } \sum_{j=0}^{n-1} \beta'_{i,j}= 0.
    $$
    Values $\beta'_{i,j}$ are random in $\mathbb{Z}_p$ conditioned on their sum being equal to $0$. 
    $\mathcal{B}$ gives 
    $$sk_{i,B_i} = \{sk_{i,j,b_{i,j}}\}_{j=0}^{n-1}
    \mbox{ and } key_i$$
to $\mathcal{A}$. Notice that from this moment onward $\mathcal{A}$ can query the PRP oracle for $key_i$ (by assumption, at most $Q_P$ times; notice that ${\cal B}$ queried the PRP oracle for $key_i$ at most two times -- if $i=i^*$ -- and the size of $Set_{PRP}(i)$ is $Q_P+2$).
\end{itemize}

\noindent
We need to verify that the $sk_{i,B_i}$ and $aux_i$ have the correct form. That is, see (\ref{eq:sk}),
$$ aux_i = g^{r_i}
\mbox{ and }
sk_{i,j,b_{i,j}} = g_2^\alpha F(it^n + b_{i,j} n t^j)^{r_i} v_{i,j}$$
for  some $\alpha$, $r_i$, and $v_{i,j}$: We will use 
\begin{eqnarray*}
\alpha &=& a, \\
r_i &=& r'_i - \frac{b}{k_i-k*}, \\
v_{i,j} &=& g_2^{\frac{-\alpha'}{k_i-k^*}-\alpha} F(it^n+b_{i,j} n t^j)^{r'_i-r_i} v'_{i,j}.
\end{eqnarray*}
Since $r'_i$ is random, also $r_i$ is random. Since all $v'_{i,j}$ are random conditioned on $\prod_{j=0}^{n-1} v'_{i,j}=1$, the definition of $v_{i,j}$ transforms $sk_{i,j,b_{i,j}}$ into the required form  if the product of all $v_{i,j}$ equals $1$. We derive
\begin{eqnarray*}
\prod_{j=0}^{n-1} v_{i,j} 
&=& \prod_{j=0}^{n-1} g_2^{\frac{-\alpha'}{k_i-k^*}-\alpha} F(it^n+b_{i,j} n t^j)^{r'_i-r_i} \\
&=& 
g_2^{\frac{-\alpha'n}{k_i-k^*}-n\alpha}
\left( \prod_{j=0}^{n-1}
g_1^{it^n+b_{i,j} n t^j}h \right)^{r'_i-r_i} \\
&=& 
g_2^{\frac{-\alpha'n}{k_i-k^*}-n\alpha}
(g_1^{(it^n+B_i)n} h^n)^{r'_i-r_i} \\
&=&
g_2^{\frac{-\alpha'n}{k_i-k^*}-na}
(g_1^{k_in} (g_1^{-k^*}g^{\alpha'})^n)^{\frac{b}{k_i-k^*}} \\
&=&
g^{\frac{-\alpha'n b}{k_i-k^*}-nab}
g^{k_ina\cdot \frac{b}{k_i-k^*}} (g^{-k_i^*na}g^{\alpha'n})^{\frac{b}{k_i-k^*}} = 1.
\end{eqnarray*}
Also 
$$aux_i = g_2^{\frac{-1}{k_i-k^*}} g^{r'_i} = g^{\frac{-b}{k_i-k^*}} g^{r'_i}=g^{r_i}$$
as required.

\begin{itemize}
    \item {\bf Guess}: Adversary $\mathcal{A}$ produces a signature $(\sigma,M^*)$ for session $\hat{i}$ such that $M^*\neq M_{\hat{i}}$.  $\mathcal{B}$ calls the PRP oracle for input $(\hat{i},M^*)$ which returns a value $\hat{B}$ from $Set_{PRP}(\hat{i})$. 
    If $(\hat{i},\hat{B})=(i^*,B^*)$, then $\mathcal{B}$ will use signature $\sigma$ in its attempt to solve CDHP.
    
\end{itemize}

\noindent
    If  $(\hat{i},\hat{B})\neq (i^*,B^*)$, then $\mathcal{B}$ repeats the same computation with the same random tape for $\mathcal{A}$, the same hash and PRP oracles, but with a fresh random chosen $(i^*,B^*)$ in Setup and fresh random choices for $v'_{i,j}$ and $r'_i$ by ${\cal B}$ in Query and Setup. Notice that even though this changes $k^*$ and as a consequence also the values for $h$ and all the key material that will be revealed to ${\cal A}$, to ${\cal A}$ it seems that all this key material is independent of $k^*$ as it matches their generation with the random looking values $v_{i,j}$ and $r_i$. Therefore, ${\cal A}$'s  choice of $(\hat{i},M^*)$ and the resulting $\hat{B}$ are independent of $k^*$. Since there are $N$ choices for $i^*$ and $Q_P+1$ choices for $B^*$ (notice that $B^*\in Set_{PRP}(i^*)\setminus \{B_{i^*}\}$ which has $Q_P+1$ elements) it will take $T(Q_P+1)N$ expected time until a usable signature $\sigma$ is computed (for which $(\hat{i},\hat{B})= (i^*,B^*)$ is a lucky guess during Setup).

$\mathcal{B}$ simulates $\mathcal{A}$ a second time for (1) the same random tape (this implies that  both the first and second simulation select the same value $s$ when producing the final signature, hence, the same value $x$ is used), (2) the same PRP oracle with the same sequence $\{key_i\}$,
but with (3) a random oracle which simulates a different hash function $H'(\cdot)$. $\mathcal{B}$ does its simulation for the same $(i^*,B^*)$ pair and we call a signature valid if it verifies properly using \Call{Verify}{} and is  usable in that $(\hat{i},\hat{B})=(i^*,B^*)$ as described above (this takes another $T(Q_P+1)N$ expected time). Since the PRP oracle defines permutations, this is equivalent to generating a signature  for $(i^*,M^*)$ that verifies properly and where $M^*$ is output by the first simulation.
Since the PRP oracle with $\{key_i\}$ is fixed, the signature scheme becomes generic in that, for $\sigma'=(x,y,z)$, values $y$ and $z$ only depend on $x$, $u=H(M^*,x)$, and secret keys, but no other randomly selected values. This allows us to apply the forking lemma.

We apply the forking lemma which states that $\mathcal{B}$ is able to use $\mathcal{A}$ to produce  two valid signatures for $(i^*,M^*)$:
$$
\sigma' = (x,y,z) \mbox{ with hash value } u=H(M,x)
$$
and
$$
\sigma" = (x,y',z') \mbox{ with hash value } u'=H'(M,x)
$$
such that $u\neq u'$ in expected time $T'\leq 84480\cdot 2T(Q_P+1)NQ_H/\epsilon$. 


Since both signatures are valid we know
\begin{eqnarray*}
e(g,z) &=& e(g_1,g_2^{nu} x y^{k^*}) \times e(y,h), \\
e(g,z') &=& e(g_1,g_2^{nu'} x y'^{k^*}) \times e(y',h).
\end{eqnarray*}
This implies that
$$ (e(g,z) \times e(g,z')^{-1})^{\frac{1}{n(u-u')}}
= e(g, (z/z')^{\frac{1}{n(u-u')}})$$
is equal to the product of
\begin{eqnarray*}
&& (e(g_1,g_2^{nu} x y^{k^*}) \times e(g_1,g_2^{nu'} x y'^{k^*})^{-1})^{\frac{1}{n(u-u')}} \\
&=&
e(g_1, g_2^{n(u-u')} (y/y')^{k^*})^{\frac{1}{n(u-u')}} \\
&=&
e(g_1,g_2 (y/y')^{\frac{k^*}{n(u-u')}}) \\
&=&
e(g_1,g_2) \times e(g_1,(y/y')^{\frac{k^*}{n(u-u')}}) 
\end{eqnarray*}
and
\begin{eqnarray*}
&& (e(y,h) \times e(y',h)^{-1})^{\frac{1}{n(u-u')}} \\
&=& 
e(y/y',h)^{\frac{1}{n(u-u')}} \\
&=&
e(y/y',g_1^{-k^*}g^{\alpha'})^{\frac{1}{n(u-u')}} \\
&=&
(
e(y/y',g_1^{-k^*}) \times e(y/y',g^{\alpha'})
)^{\frac{1}{n(u-u')}} \\
&=&
e((y/y')^{\frac{-k^*}{n(u-u')}},g_1) \times
e((y/y')^{\frac{\alpha'}{n(u-u')}},g).
\end{eqnarray*}
This proves 
\begin{eqnarray*}
&& e(g, (z/z')^{\frac{1}{n(u-u')}}) \\
&=& 
e(g_1,g_2) \times e(g_1,(y/y')^{\frac{k^*}{n(u-u')}})  \\
&& \times e((y/y')^{\frac{-k^*}{n(u-u')}},g_1) \times
e((y/y')^{\frac{\alpha'}{n(u-u')}},g)\\
&=& 
e(g_1,g_2) \times e(g,(y/y')^{\frac{\alpha'}{n(u-u')}}).
\end{eqnarray*}
By using $g_1=g^a$ and $g_2=g^b$ we obtain
$$
e(g, ((z/z')/(y/y')^{\alpha'})^{\frac{1}{n(u-u')}})= e(g,g^{ab})
$$
from which we conclude
$$((z/z')/(y/y')^{\alpha'})^{\frac{1}{n(u-u')}} = g^{ab}.$$
This means that the two valid  signatures allow $\mathcal{B}$ to compute $g^{ab}$ and this solves CDHP. We notice that the above reduction, which only uses the validity checks of the two signatures, is not present in~\cite{ChowHuiYiuEtAl2004}; the above analysis can be adapted to~\cite{ChowHuiYiuEtAl2004} in order to properly complete their security analysis.  

\vspace{3mm}

\noindent
{\bf Checking $s+u\neq 0$.}  The signing procedure of~\cite{ChowHuiYiuEtAl2004} repeats selecting a random $s$ and computing $u=H(M,x)$ until $s+u\neq 0$. 
When verifying we may check $g_2^{nu}x \neq 1$ as this corresponds to  $s+u\neq 0$.
We notice that in our security analysis we do not need this additional property $s+u\neq 0$. But in practice we may exclude $s+u=0$ since this is easy to do and if not excluded, then any one who happens to solve $s+H(M,g_2^{ns})=0$ (by using some table enumeration method) will immediately be able to use this to impersonate a corresponding signature (for $s+u=0$, $y=z=1$). 

\subsection{Compressed Signatures}
\label{sec:opt}

To reduce the computational overhead of the signer, we can directly forward the selected secret keys and the corresponding auxiliary information to the verifier without completing the signing procedure. The secret key generation procedure remains the same in the scheme, so the security proof of the above scheme still holds. The signing and verification procedures are described below.

\noindent
{\bf {Signing.}} 
We compute $B=\sum_{j=0}^{n-1} b_jt^j$ with $0\leq b_j<t$ as the pseudo random permutation output
$B = \Call{PRP}{key;M}$ for a random $key$. We define  subset $\phi(B)=\{tj+b_j\}_{j=0}^{n-1}$ of $\{0,\ldots, q-1\}$ for $q=tn$. We represent its elements by the pairs $(j,b_j)$.
We produce  
$$ \sigma' \leftarrow \Call{Sign'}{\{sk_{i,j,b_j}\}_{j=0}^{n-1},aux_i;M},$$
which signs a plaintext $M \in \mathbb{G}_1$ 
as follows:
\begin{itemize}
\item Compute 
\begin{eqnarray*}
sk_i &=& \prod_{j=0}^{n-1} sk_{i,j,b_j} 
= \prod_{j=0}^{n-1} 
g^{\alpha}_2 F(it^n+b_jnt^j)^{r_i}v_{i,j} \\
&=&
g^{n\alpha}_2 \left( \prod_{j=0}^{n-1} F(it^n+b_jnt^j) \right)^{r_i}  \\
&=&
g^{n\alpha}_2 (g_1^{it^n+B}h)^{nr_i} 
=
(g_2^\alpha F(it^n+B)^{r_i})^n. 
\end{eqnarray*}
\item Return signature $\sigma=(\sigma',key)$ for \begin{eqnarray*}\sigma' &=& (y,z) \\
&=& (aux_i,sk_i) \\
&=& (g^{r_i},(g^\alpha_2F(it^n+B)^{r_i})^{n} ).
\end{eqnarray*}
\end{itemize}

\vspace{3mm}

\noindent
{\bf Verification.} \Call{Verify}{$pk,i;\sigma,M$} with $\sigma=(\sigma',key)$  verifies   signature $\sigma'=(y,z)$ for message $M$, where $\sigma'$ is generated during the $i$-th session:
\begin{itemize}
\item Compute $B=\Call{PRP}{key;M}$.
\item The signature verifies  if and only if 
$$e(g,z)=e(g_1,g^{n}_2) \times e(y,(g_1^k h)^n) \mbox{ with } k= it^n+B.$$ 
\end{itemize}

\noindent{\bf Correctness.}
The correctness of the scheme follows from
$g_1= g^\alpha$, $y = g^{r_i}$, $z=(g^\alpha_2F(k)^{r_i})^{n}$, and 
\begin{eqnarray*}
&& e(g_1, g_2^n) \times e (y, (g_1^kh)^n) \\
&=& e(g^\alpha,g^{n}_2) \times e(g^{r_i},(g_1^kh)^n) \\
&=& e (g^\alpha, g_2)^n \times e(g^{r_i}, g_1^kh)^n
\\
&=& (e (g^\alpha, g_2) \times e(g^{r_i}, g_1^kh))^n
\\
&=& (e (g, g_2^\alpha) \times e(g, (g_1^kh)^{r_i}))^n
\\
&=& (e (g, g_2^\alpha (g_1^kh)^{r_i}))^n
\\
&=& (e (g, g_2^\alpha F(k)^{r_i}))^n
\\
&=& e (g, (g_2^\alpha F(k)^{r_i})^n)
\\
&=& e(g,z).
\end{eqnarray*}

\section{Pseudo-codes for OTS-SKE Implementation}
\label{sec:pseudocodes}
In this section, we present the pseudo-codes of the bilinear-map based OTS-SKE that is presented in Appendix \ref{sec:bilinear} and used in our performance evaluation. Algorithm \ref{alg:bilinear_pseudocode} describes the implementation of our Bilinear-map based OTS-SKE ${\cal S}=(\Call{KeyGen}{}, \Call{Sign}{}, \Call{Verify}{})$.\\

Given a session $i$, \Call{KeyGen}{} generates auxiliary information $aux_i$ and a set of secret keys $sk_i = \{sk_{i,j,b}\}_{j=0,b=0}^{n-1,t-1}$. Using algorithm $IG$ and security parameter $\alpha$, a set of parameters $(p,\mathbb{G},\mathbb{G}_1,e,g,g_2)$ are generated and then combined with $g_1$ and $h$ that are randomly generated to produce a public key $pk$ (Line 1--5). After the public key generation, $aux_i$ and a total of $tn$ secret keys are generated of which a subset will be extracted from using a random nonce during signing (Line 6--13). In the implementation that was presented in Section \ref{sec:performance}, the constant components of the secret keys (given in Line 11) are precomputed before generating the keys for optimization. \\

\Call{Sign}{} takes a session id $i$, a random nonce and a message $M$, and computes $B$ using a pseudo random permutation \Call{PRP}{nonce;M} where $B=\sum_{j=0}^{n-1} b_jt^j$ (Line 18). Using this string $B$, a set of $n$ secret keys out of $tn$ keys are extracted and combined (Line 19--21). Finally, the auxiliary information along with the final product of the secret keys are combined $\sigma' =  (aux_i,sk_{prod})$ and the signature $\sigma = (\sigma', nonce)$ is returned (Line 22--23). Note that this signing process is completed by computing a string $B$ based on the user's received random nonce and the message $M$ and extracting a set of secret keys unique to $B$. For this reason, the keys and the auxiliary information can be directly forwarded to the user as a signature, rather than performing an actual signing operation on message $M$. This reduces the computational overhead of our signing process, as it avoids 
complex and costly elliptic curve operations such as pairings.\\

\Call{Verify}{} receives the public key $pk$, session id $i$, the signature $\sigma$ and the message $M$. It extracts the parameters of $pk$, along with $\sigma' =  (aux_i, sk_{prod})$ and $nonce$ from the signature $\sigma$ (Line 27--30). Using the extracted parameters, 3 pairings on the elliptic curve $e_1$, $e_2$ and $e_3$ are performed (Line 31--34). \Call{Verify}{} returns True only and only if $e_1 = e_2\times e_3$. Otherwise, the verification of the signature fails (Line 35--39). Note that the verification process ($127.3$ ms) is significantly more costly than the signing phase ($3.4$ ms) and the baseline method Elliptic Curve Digital Signature Algorithm (ECDSA) verification ($74$ ms). This is because multiple pairings on the elliptic curve are performed during verification, whereas this cost is avoided during signing by making the keys unique to a random nonce and a message. We refer the readers to Section \ref{sec:performance} for a more detailed analysis of the performance and the comparison with the baseline.

\begin{algorithm}[tbh!]
\caption{Bilinear-map based OTS-SKE}\label{alg:bilinear_pseudocode}

We assume a OTP-SKE-EUF-CMA secure bilinear-map based OTS-SKE scheme ${\cal S}=(\Call{KeyGen}{}, \Call{Sign}{}, \Call{Verify}{})$. Algorithm $IG$ generates a suitable mathematical structure for our signature scheme and $\lambda$ is the security parameter. We assume a Pseudo Random Number Generator (PRNG) bootstrapped from an initial seed extracted from a True Random Number Generator (TRNG) to generate random bit strings.

\begin{algorithmic}[1]
\Procedure{KeyGen}{session $i$}

\State $(p,\mathbb{G},\mathbb{G}_1,e,g,g_2)\leftarrow IG(1^\lambda)$
\State $(\alpha,h) \leftarrow$  PRNG
\State $g_1=g^{\alpha}$
\State $pk \gets (p,\mathbb{G},\mathbb{G}_1,e,g,g_1,g_2,h)$ 

\State $r_i \gets $ PRNG
\State $aux_i \gets g^{r_i}$
\State Generate $V$ \Comment{ $\prod_{j=0}^{n-1} V_{j} = 1$} 
\For{$j \gets 0$ to $n-1$} 
\For{$b \gets 0$ to $t-1$}
\State $sk_{i,j,b}=g^{\alpha}_2 (g_1^{it^n+bnt^j} h)^{r_i}V_{j}$
\EndFor
\EndFor
\EndProcedure\\

\Procedure{Sign}{session $i$, $nonce$, $M$}
\State Initialize $sk_{prod} \gets 0$
\State $B = \Call{PRP}{nonce;M}$ \Comment{$B=\sum_{j=0}^{n-1} b_jt^j$}
\For{$j \gets 0$ to $n-1$}
\State $sk_{prod} = sk_{prod} \times sk_{i,j,b_j}$ 
\EndFor
\State $\sigma' \gets (aux_i,sk_{prod})$
\State \Return $\sigma \gets (\sigma', nonce)$ 
\EndProcedure\\

\Procedure{Verify}{$pk$, session $i$, $\sigma$, $M$}

\State $(p,\mathbb{G},\mathbb{G}_1,e,g,g_1,g_2,h) \gets pk $
\State $(\sigma', nonce) \gets \sigma$
\State $B=\Call{PRP}{nonce;M}$
\State $(aux_i,sk_{prod}) \gets \sigma'$
\State $e_1 = e(g,sk_{prod}) $
\State $e_2 = e(g_1,g^{n}_2) $
\State $k= it^n+B$
\State $e_3 = e(aux_i,(g_1^k h)^n)$
\If {$e_1 == e_2\times e_3$}
\State \Return True
\Else
\State \Return False
\EndIf
\EndProcedure

\end{algorithmic}

\end{algorithm}
 
\clearpage
\newpage

\bibliographystyle{IEEEtran}
\bibliography{references}

\end{document}